\documentclass[fleqn,usenatbib]{mnras}

\usepackage{newtxtext,newtxmath}

\usepackage[T1]{fontenc}
\usepackage{ae,aecompl}
\usepackage{booktabs}
\usepackage{multirow}
\usepackage{siunitx}
\usepackage{rotating}

\usepackage{graphicx}	
\usepackage{float}
\usepackage{amsmath}	
\usepackage{newtxtext,newtxmath}
\usepackage{float}
\usepackage{ulem}

\title[The proper motion of stars in dwarf galaxies]{The proper motion of stars in dwarf galaxies: distinguishing central density cusps from cores}

\author[I. de Martino et al.]{
Ivan de Martino$^{1,2}$\thanks{E-mail: ivan.demartino@usal.es}, Antonaldo Diaferio$^{1,3}$, Luisa Ostorero$^{1,3}$ 
\\
$^{1}$ Dipartimento di Fisica, Universit\`a di Torino,  Via P. Giuria 1, I-10125 Torino, Italy\\
$^{2}$  Universidad de Salamanca,Facultad de Ciencias.,F\'isica Te\'orica, Salamanca, Plaza de la Merced s/n. 37008, Spain\\
$^{3}$  Istituto Nazionale di Fisica Nucleare (INFN), Sezione di Torino, Via P. Giuria 1, I-10125 Torino, Italy
}

\date{Accepted XXX. Received YYY; in original form ZZZ}

\pubyear{2021}

\begin{document}
\label{firstpage}
\pagerange{\pageref{firstpage}--\pageref{lastpage}}
\maketitle

\begin{abstract}
We show that measuring the proper motion of ${{\sim 2000}}$ stars within a dwarf galaxy, with an uncertainty of 1 km/s at most, can establish whether the Dark Matter (DM)  density profile of the dwarf has a central core or cusp. We derive these limits by building mock star catalogues similar to those expected from future astrometric {\it Theia}-like missions  and including celestial coordinates, radial velocity and proper motion of the stars. The density field of the DM halo of the dwarf is sampled from an extended Navarro-Frank-White (eNWF) spherical model, whereas the number density distribution of the stars is a Plummer sphere. The velocity field of the stars is set according to the Jeans equations. A Monte Carlo Markov Chain algorithm applied to a sample of $N\gtrsim 2000$ stars returns unbiased estimates of the eNFW DM parameters within $10\%$ of the true values and with $1\sigma$ relative uncertainties $\lesssim 20$\%. The proper motions of the stars lift the degeneracy among the eNFW parameters which appears when the line-of-sight velocities alone are available. {Our analysis demonstrates that, by estimating the log-slope of the mass density profile estimated at the half-light radius, a sample of $N=2000$ stars can distinguish between a core and a cusp at more than $8\sigma$.} Proper motions also return unbiased estimates of the dwarf mass profile with $1\sigma$ uncertainties that decrease, on average, from 2.65 dex to  0.15 dex when the size of the star sample increases from $N=100$ to $N=6000$ stars. The measure of the proper motions can thus strongly constrain the distribution of DM in nearby dwarfs and provides fundamental contribution to understanding the nature and the properties of DM.

\end{abstract}



\section{Introduction}

Indirect observational pieces of evidence suggesting the existence of  dark matter (DM) have been accumulating over the last decades (see, e.g., \citealt{Salucci_2018}, and references therein). 
DM is suggested to be made of particles emerging from the supersymmetric extensions of the standard model \citep{Ellis1984}; these hypothetical particles are weakly interacting with standard matter, they are non-relativistic and basically collisionless (for 
comprehensive reviews  see, e.g.,  \citealt{Bertone2005,Feng2010,Frenk2012,Primack2012,Freese2017,Bertone2018}). 
This Cold Dark Matter (CDM) paradigm successfully explains the evolution and formation of the large-scale structure \citep[e.g.,][]{Planck2018}. 
However, a direct detection of dark matter particles 
is still missing \citep{Akerib2017,Aprile2018,Tanabashi2018}. In addition, the CDM model shows some tensions with observations both at cosmological scales (\citealt{Ade2016,Riess2016,Schwarz2016, Riess2019, DiValentino2020, Mathews2020,Handley2021,Luongo2021})
and at galactic scales  (\citealt{Weinberg2015,DelPopolo2017,Bullock2017,idm2020b}).

Among the latter tensions, the core-cusp problem (CCP)
is the discrepancy between the inner slope of the DM  density profile of galaxies, inferred from observations, and the profile that is  predicted by DM-only N-body simulations  
\citep[e.g.,][]{Navarro1997}: large cores appear to be commonly present in real dwarf galaxies (\citealt{Carignan1988,Carignan1989,deBlok1997,Mateo1998,deBlok1997,Gentile_2007,Oh_2015,Torrealba2016,Caldwell2017,Torrealba2019}), whereas N-body simulations predict that DM haloes should have a central density cusp. Although  hydrodynamical simulations, that include dissipative processes, suggest that energy feedback from stars, supernovae explosions, and high-energy phenomena may turn the cusp of the DM profile into a core 
\citep{Mashchenko2008,Pontzen2012,Governato2012,Onorbe2015,Read2016}, 
more recent hydrodynamical simulations seem to disfavour this and related mechanisms  \citep{Benitez-Llambay2019,Bose2019}. 

A more drastic solution might require a change in the CDM paradigm. This change may involve either the nature of the dark matter particles, for example the replacement of the CDM particles with self-interacting dark matter \citep{Rocha2013} or ultra-light axions \citep{Chen2017,Broadhurst2019,Pozo2020}, or a  modification of the theory of gravity \citep[e.g.,][]{Famaey2012,idm2016,idm2017, DelPopolo2017,idm2020}. 

Self-interacting DM, rather than CDM, can provide a solution to the CCP by generating a central core without resorting to baryonic feedback 
\citep[e.g.][]{Spergel2000, Rocha2013}. A similar solution can be provided by fuzzy DM, where the DM particles are very light bosons with mass $\sim10^{-22}$ eV \citep{Hu2000,Chen2017,Broadhurst2019,idm2020PDU,Pozo2020}. 
\citet{Khoury2015} proposed that  DM made of axion-like particles can actually be a superfluid on galactic scales and keep the collisionless behaviour of standard CDM on larger scales. 
The additional force induced by this superfluid DM on galactic scales appears to be consistent with the observed phenomenology of galaxies, some of which appears problematic for the standard CDM model but was largely predicted by  Modified Newtonian Dynamics (MOND)  \citep[][]{Milgrom1983,Bekenstein1984,Famaey2012,McGaugh2020}. 

On the observational side, the presence of cores rather than cusps is still disputed \citep{van_den_Bosch_2001,Spekkens_2005,Walker2009,DC14a,Marsh2015}. Indeed,  
current data appear unable to unambiguously distinguish between a central cusp and a core in dark matter density profiles (e.g., \citealt{Walker2009}). Although the recent discoveries of ultra-faint galaxies seem to suggest the existence of a large core in their innermost region on the basis of their velocity dispersion profile (see for instance \citealt{Torrealba2019}), the constraining power of these observations is weak and the debate on whether dwarf galaxies do or do not exhibit a large dark matter core is still open.  

The method to map the distribution of DM depends on the type of system under investigation: 
for a rotationally-supported galaxy, the DM distribution is generally inferred 
from the fit to the rotation curve, whereas, for a pressure-supported dwarf spheroidal galaxy, we rely on the profile of the line-of-sight velocity dispersion, if no other data set, like  multiple stellar populations \citep[e.g.,][]{Walker2009,Walker2011}, proper motions \citep[e.g.,][]{Strigari2007,Evslin2015}, or three dimensional positions \citep[e.g.,][]{Richardson2014} is available. 

The standard approach is  to assume a  functional form for the DM halo density profile and determine its parameters from a fit to the data. Unfortunately, 
this approach generally suffers from degeneracies among the parameters of the DM density profile and even among them and other unknown parameters (\citealt{Lokas2002,DeLorenzi2009,Walker2009,Napolitano2014, Read2017}).  A well-known case is the degeneracy between the velocity anisotropy parameter $\beta$ and the total halo mass in the Jeans equations \citep[e.g.,][]{Lokas2002,Binney2008,Campbell_2017}. 
This drawback may inhibit the distinction between 
models with a cusp and with a core, as it happens when fitting
the line-of-sight velocity dispersion profiles of
the Milky Way satellites \citep{Walker2009,2013A&A...558A..35B,2013arXiv1305.0670R}. 
This degeneracy can be lifted by adding information from multiple stellar populations \citep[e.g.,][]{Walker2011,Zhu2016a,Zhu2016b} or higher velocity moments \citep[e.g.,][]{Lokas2003}. However, N-body simulations show that multiple stellar populations only partially lift the degeneracy, whereas  higher velocity moments combined with proper motions appear to be more effective \citep{Read2017,Webb2018}.  
\citet{2018NatAs...2..156M,massari2020}
use the proper motions of only a few hundred stars 
to estimate the velocity anisotropy parameter 
$\beta=0.86^{+0.12}_{-0.83}$ and $\beta=0.25^{+0.47}_{-1.38}$ in the Sculptor and Draco dwarf galaxies, respectively.  

Here, we quantify how proper motions can indeed lift the mass-anisotropy degeneracy and shed light on the CCP in dwarf galaxies. The proper motions of the stars of the dwarf combined with their line-of-sight velocities from their spectra provide the  three-dimensional velocity field within the dwarf. \citet{Strigari2007} pointed out that adding the proper motion  of 200 stars to their line-of-sight velocity would  make it possible to constrain the log-slope of the DM density profile at twice the King radius with 20\%  statistical uncertainty, whereas using only the line-of-sight velocities leaves the log-slope parameter unconstrained. 

More recently, \citet{Watkins2013} (hereafter W13) suggested a procedure to model proper motion data sets in axisymmetric systems, and applied their approach to the star cluster $\omega$-Centauri.\footnote{{ By adopting the approach of \citet{Watkins2013}, \citet{Evans2022} use the most recent kinematic data of $\omega$-Centauri to constrain both the distribution of the non-luminous mass component of this globular cluster and, if this dark component is made of elementary particles, the J-factor associated with the observable flux of $\gamma$-rays coming from dark matter particle annihilations.  }} By assuming that the velocity distribution of the stars in the system is described by a trivariate Gaussian, W13 constructed a probability distribution function of the 3-dimensional velocity field based on the solution of the Jeans equations that can be applied to discrete velocity data. 
In this technique, \citet{Zhu2016a}  included a chemo-dynamical model with multiple stellar populations.
\citet{Zhu2016b} (hereafter Z16) applied this combined technique to a star catalogue of the Sculptor dwarf spheroidal; they use mock data to estimate that a dataset of approximately 6000 stars with line-of-sight velocities and metallicity information but no proper motions  is required to solve the CCP.

Here, we assume a single population of stars with no metallicity information and perfectly known celestial coordinates and adopt the approach of W13 to estimate the parameters of the DM density profile based on kinematical information alone. We create mock data sets mirroring the generic expected observational limitations of future space-borne astrometric missions to determine the minimum number of stars and the maximum uncertainty on the proper motions that are required to  effectively solve the CCP. As a benchmark, we adopt a future {\it Theia}-like mission \citep{Malbet2016,Theia2017,Malbet2019,Malbet2021}
designed to achieve an end-of-mission uncertainty on proper motions of a few $\mu$as~yr$^{-1}$, namely $\sim 100$ times smaller than the uncertainty of {\it Gaia} \citep{Gaia2016,Gaiaetal2016a,Gaiaetal2016b,Gaia2018,Gaiaetal2018,Gaiaetal2021}. A {\it Theia}-like mission is thus expected to be able to measure the proper motions of stars in nearby dwarf galaxies, and thus solve the CCP  through an accurate determination of the DM density profile. 
{ A solution to the CCP based on proper motions may also be provided by future 30 meter class telescopes as, e.g., the Thirty Meter Telescope (TMT)\footnote{\tt \url{http://www.tmt.org}} \citep{Schoeck2013,Skidmore2015}, or by a combination of measures from these telescopes and space-based facilities \citep{Evslin2015}.}

Section \S \ref{sec:methodology} describes the mock catalogues and Section \S \ref{sec:dataprocessing}  our  methodology for the data processing. Section \S \ref{sec:results_sec} shows  how our procedure is effective in constraining the model parameters and  solving the CCP. Section \S \ref{sec:CCP&Mass} explores the dependence of the accuracy of the estimated cumulative mass profile of the dwarf on the size of the star sample and on the knowledge of proper motions. We summarize our results and conclude in Section \S \ref{sec:discussion}.

\section{Method}\label{sec:methodology}

The creation of an astrometric mock catalogue of the stars in a dwarf galaxy requires the assumption of the phase-space  distributions of both the stellar and the DM components, and the sampling of the stellar distribution.  We adopt two assumptions: (1) both the DM and the star distributions are spherically symmetric; (2) the anisotropy velocity parameter $\beta$ is independent of radius.

\subsection{Stellar Distribution}

Following \citet{Walker2009}, for the stellar number density distribution we adopt the \citet{Plummer1911} model
\begin{equation}\label{nup}
\nu(r) \propto \left(  1 + \frac{r^{2}}{a^{2}} \right)^{-5/2},
\end{equation}
where $a$ is a scale length. The stellar mass density is proportional to 
$\nu(r)$. 
In dwarf spheroidal galaxies, DM dynamically dominates over the stellar mass density component at any radius, including the central regions \citep[e.g.,][and references therein]{idm2020b}; therefore, the stellar contribution to the total gravitational dynamics is negligible. Hereafter, we thus ignore the contribution of the stars to the dynamics of the dwarf. 

\subsection{Dark matter distribution}

We model the DM mass density distribution  with the profile  \citep{Hernquist1990, Zhao1996}
\begin{equation}\label{DMdensityEqn}
\rho(r) = \rho_{0} \left(\frac{r}{r_{s}}  \right)^{-\gamma} \left[ 1 + \left( \frac{r}{r_{s}}  \right)^{\alpha} \right]^{\frac{\gamma - \delta}{\alpha}}\, .
\end{equation}
The free parameters are the central density $\rho_0$, the scale radius $r_s$, and the three slope parameters $\alpha$, $\delta$, and $\gamma$.\footnote{The parameter $\delta$ is usually called $\beta$. Here, we adopt a different notation to avoid confusion with the velocity anisotropy parameter $\beta$ appearing in the Jeans equations.}  The value $\gamma=0$ corresponds to a density profile with a central core, whereas $\gamma>0$ indicates a profile with a cusp. Equation (\ref{DMdensityEqn})  defines an extended  Navarro-Frenk-White (eNFW) profile and reduces to the standard Navarro-Frenk-White profile \citep{Navarro1997} when $(\alpha,\delta,\gamma)=(1,3,1)$. In our models, we always set $\alpha=1$ and $\delta=3$. We explore the  profiles with a core and with a cusp by setting $\gamma=0$ and $\gamma=1$, respectively, similarly to \citet{Walker2009}.

With the density profile of Eq.~\eqref{DMdensityEqn}, the cumulative mass within  radius $r$ is
\begin{equation}
M(<r) = \frac{4 \pi \rho_0 r_s^3}{3-\gamma}\left( \frac{r}{r_s} \right)^{3-\gamma} {}_2 F_1 \left(3-\gamma,3-\gamma ;3-\gamma+1;-\frac{r}{r_s}\right),
\label{eq:MassNFW}
\end{equation}
 where ${}_2F_1$ is the hypergeometric function. 

\subsection{Sampling the phase-space distributions}
\label{sec:SamplingPhaseSpace}

We consider a sample of $N$ stars with spherical coordinates $(r,\theta,\phi)$ in the reference frame centered on the centre of mass of the dwarf: the  radius $r$ derives from sampling the Plummer density profile of Eq.~\eqref{nup}, whereas the angular coordinates are sampled from  uniform distributions in the ranges $\cos \theta = (-1,1)$ and $\phi = (0,2\pi)$. 

We sample the three velocity components of each star from the velocity distribution function, assumed to be a multivariate Gaussian distribution. The probability of finding a velocity ${ v}$ at position ${\boldsymbol r}$ is thus
\begin{equation}\label{Likelihood1}
p({ v}| {\boldsymbol r}) = \frac{\exp \left\{ -\frac{1}{2}\left[ { v} - {\boldsymbol \mu({\boldsymbol x})} \right]^{T} { C^{-1}({\boldsymbol r})}\left[ { v} - {\boldsymbol \mu({\boldsymbol r})} \right]\right\} }{\sqrt{\left( 2\pi \right)^3\left| { C({ r})}\right|}}\, ,
\end{equation} 
where 
\begin{equation}
{\boldsymbol \mu}({\boldsymbol r}) = \begin{pmatrix}
\overline{{{v}}_{x}} \\
\overline{{{v}}_{y}} \\
\overline{{{v}}_{z}} \\
\end{pmatrix}
\end{equation}
is the mean velocity at position ${\boldsymbol r}$, and 
\begin{equation}\label{eq:cov}
{ C}({\boldsymbol r}) = \begin{pmatrix}
\overline{{{v}}_{x}^{2}} - \overline{{{v}}_{x}}^{2} & \overline{v_{xy}^{2}} - \overline{{{v}}_{x}} ~\overline{{{v}}_{y}} & \overline{v_{xz}^{2}} - \overline{{{v}}_{x}} ~\overline{{{v}}_{z}}\\
 \overline{v_{xy}^{2}} - \overline{{{v}}_{x}} ~\overline{{{v}}_{y}} & \overline{{{v}}_{y}^{2}} - \overline{{{v}}_{y}}^{2} & \overline{v_{yz}^{2}} - \overline{{{v}}_{y}} ~\overline{{{v}}_{z}} \\
 \overline{v_{xz}^{2}} - \overline{{{v}}_{x}} ~\overline{{{v}}_{z}}& \overline{v_{yz}^{2}} - \overline{{{v}}_{y}} ~\overline{{{v}}_{z}}& \overline{{{v}}_{z}^{2}} - \overline{{{v}}_{z}}^{2}
\end{pmatrix}
\end{equation}
is the covariance matrix at the same position ${\boldsymbol r}$.
In the above equations, the $x$, $y$, and $z$ subscripts denote the Cartesian coordinates in the reference frame  
of the dwarf. The $x$ and $y$ axes lie on the plane of the sky, while the $z$ axis is set along the line of sight. The quantities  $\overline{v_x}$ and $\overline{v_x^2}$ are the first and second moments of the velocity distribution in the  direction of the $x$ axis, and  a similar convention is adopted for the other moments and directions. 
The equations of transformation of the star velocity from the Cartesian coordinate system, as shown in Eqns.~\eqref{Likelihood1}-\eqref{eq:cov}, to the spherical coordinate system 
are provided in Appendix \ref{velocitymoments}. 

In spherical symmetry, we have 
$\overline{v_r}=\overline{{{v}}_{\theta}}=\overline{{{v}}_{\phi}} = 0$ and $\overline{{{v}}_{r} {\rm{v}}_{\theta}}=\overline{{{v}}_{r} {\rm{v}}_{\phi}} =\overline{{{v}}_{\theta} {\rm{v}}_{\phi}} = 0 $, and 
the radial component of the Jeans equation reads
\begin{equation}
\frac{1}{\nu(r)}{{\rm d} [\nu(r) \overline{{{v}}_{r}^2}(r)]\over {\rm d} r}  + 2\beta(r) {\overline{{{v}}_{r}^2}(r) \over r}= 
- {{\rm d}\Phi(r)\over {\rm d}r}\, ,
\label{eq:jeans}
\end{equation}
with $\Phi(r)$ the gravitational potential. 
We further simplify our model by assuming that the velocity anisotropy parameter 
\begin{equation}
\beta(r) = 1 - \frac{\overline{{{v}}_{\theta}^2}(r) + \overline{{{v}}_{\phi}^2}(r)}{2\overline{{{v}}_{r}^2}(r)} =  1 - \frac{\overline{{{v}}_{t}^2}(r)}{\overline{{{v}}_{r}^2}(r)}
\label{eq:beta}
\end{equation}
is independent of $r$. In the equation above, 
$\overline{{{v}}_{\theta}^2}(r)$ = $\overline{{{v}}_{\phi}^2}(r)$ and $\overline{{{v}}_{t}^2}(r) \equiv [\overline{{{v}}_{\theta}^2}(r) + \overline{{{v}}_{\phi}^2}(r)]/2$.  
The general solution of the Jeans equation is thus
\begin{equation}
\overline{{{v}}_{r}^2}(r) = \frac{1}{\nu(r) r^{2\beta}} \int^{\infty}_{r} dr' r'^{2\beta} \nu(r') \frac{d\Phi}{dr'} \, .
\label{eq:jeanssol}
\end{equation}
Given $\nu(r)$ and $\rho(r)$ from Eqs. (\ref{nup}) and (\ref{DMdensityEqn}) and for a chosen value of $\beta$, Eq.~\eqref{eq:jeanssol} returns the covariance matrix ${ C}({\boldsymbol r})$ in Eq.~\eqref{eq:cov}
once converting from spherical to Cartesian coordinates. We can now sample the three velocity components from the velocity distribution function in  Eq.~\eqref{Likelihood1}.

\subsection{Test of the sampling}

\begin{figure}
\includegraphics[width=\columnwidth]{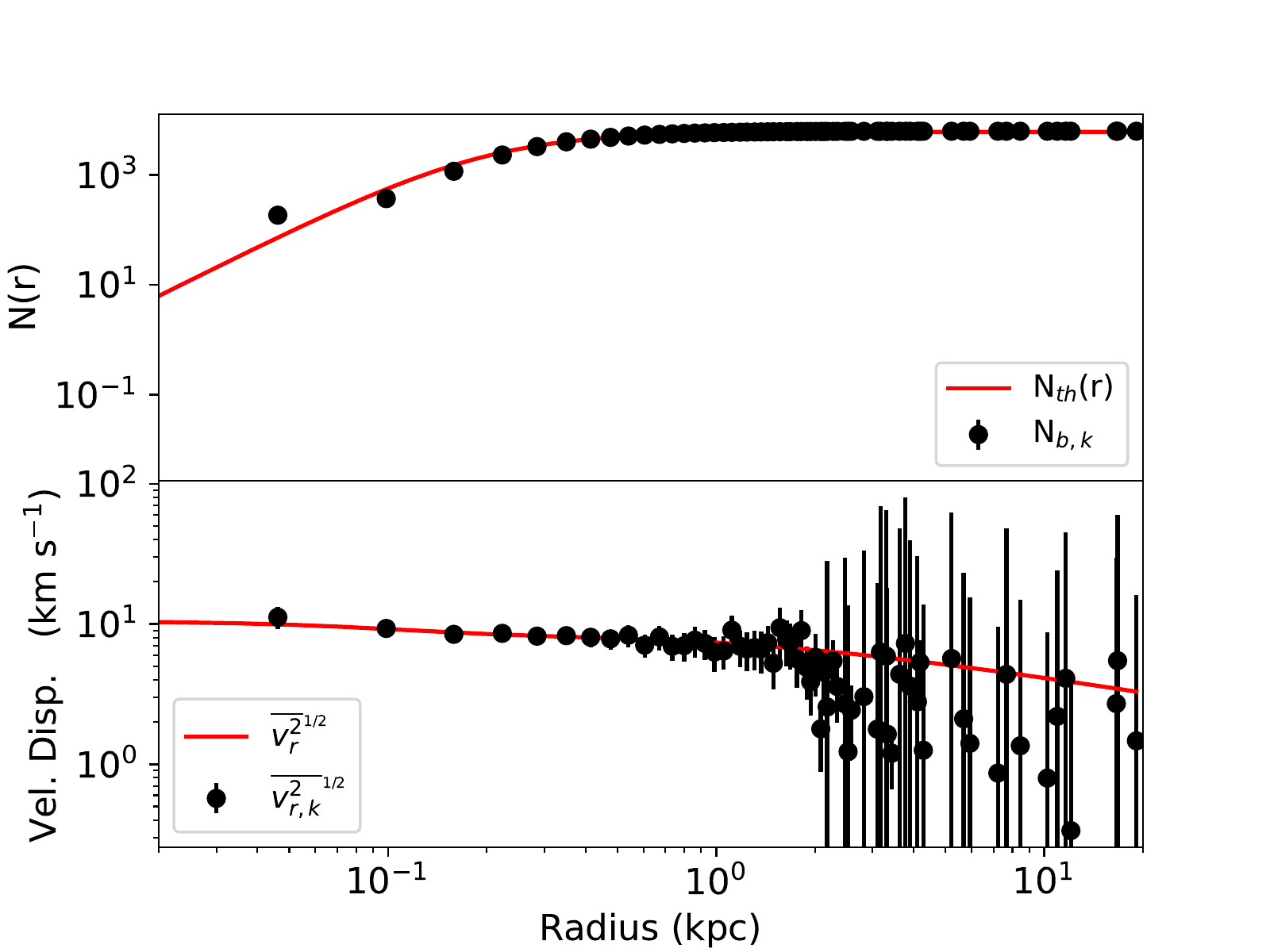} 
\caption{Example of the sampling of a dwarf with a core and isotropic velocity field $\beta=0$ with a mock catalogue with 6000 stars.  {\it Upper panel}: The cumulative number of stars in the sample (black points) and the number expected from a Plummer sphere according to Eq.~\eqref{Number} (red solid line).  {\it { Lower} panel}: The radial velocity dispersion profile according to Eq.~\eqref{eq:jeanssol} (red solid line) and Eq.~\eqref{JeansDiscrete} (black points).   
}
\label{CodeTests}
\end{figure}

To test our procedure to create the simulated dwarf, we compare the
cumulative number of stars of our sample with the cumulative number expected from a Plummer sphere 
\begin{equation}\label{Number}
N_{\rm th}(<r) = \frac{N r^3}{(r^2 + a^2)^{3/2}}\, ,
\end{equation} 
where $N$ is the total number of stars in the sample. The upper panel of Fig. \ref{CodeTests} shows the comparison for a sample of $N=6000$ stars in a DM density profile with a core, $\gamma=0$, and isotropic velocity field, $\beta=0$.   We adopt equally-spaced radial bins, each containing $N_{b,k}$ stars. The statistical fluctuation on the observed number of stars in the $k$-th bin is
\begin{equation}
\sigma_{N_{b,k}}   = N_{b,k}^{1/2}\, .
\end{equation} 
The corresponding error bars are not visible in the figure because they are smaller than the symbol size.
The agreement between our sampling and the expected profile is within the fluctuations due to shot noise. Samples with smaller $N$ show similar results.

The lower panel  of Fig.~\ref{CodeTests} compares  
the expected radial velocity dispersion profile [see Eq.~\eqref{eq:jeanssol}] with the velocity dispersion of the same sample of $N=6000$ stars. By associating each  star with radial velocity $(v_r)_i$ to the proper $k$-th radial bin, 
the radial velocity dispersion profile is
\begin{equation}\label{JeansDiscrete}
\overline{{{v}}_{r,k}^2(r)}^{1/2} = \left[\sum_i^{N_{b,k}} (v_r)_i^2/N_{b,k}\right]^{1/2}\, .
\end{equation}
The statistical fluctuation on the velocity dispersion profile in each bin is  
\begin{equation}
        \sigma_{\overline{{{v}}_{r,k}^2}^{1/2}}= \left[\frac{\overline{{{v}}_{r,k}^2}}{2(N_{b,k}-1)}\right]^{1/2}\,.
\end{equation}
Similarly to the density profile, the agreement between our sampling and the expected profile is within the fluctuations due to shot noise and supports the validity of our sampling procedure.

\begin{table}
\caption{Parameters of the mock stellar catalogues. 
The three listed values of $\beta$ and $\sigma_v$ are adopted for both the core and cusp models.}
\centering
\begin{tabular}{ccccc}
\hline
\multicolumn{2}{c}{Parameters}       & \multicolumn{2}{c} {Values}         & Unit \\
\hline
Name   & Description                & Core             & Cusp          &      \\
\hline
$\rho_{0}$ & DM central density  & 192.42           & 5.38          &  $10^{7}$ M$_{\odot}$ kpc$^{-3}$     \\
$ r_{s}$   & DM scale radius   & 150              & 795           &   pc   \\
$\alpha $ & {DM slope} & 1            & 1     \\
$\delta $ & {DM slope} & 3            & 3     \\
$\gamma$  & DM slope & 0                & 1             &      \\
$a$      & stellar scale radius       & 196 & 196        & pc  \\
$\beta$   & velocity anisotropy       & \multicolumn{2}{c}{\{-0.25, 0, 0.25\}} &      \\
$\sigma_v$ & velocity uncertainty & \multicolumn{2}{c}{\{0.0, 1.0, 5.0\}} & km s$^{-1}$ \\
\hline
\end{tabular}
\label{ParamsTable}
\end{table}

\subsection{Astrometric mock catalogues}
\label{sec:astro_mocks}
We assume that, for each star in the dwarf, the astrometric catalogue lists the two celestial coordinates, the velocity component along the line of sight, and the two components of the proper motion. With the known distance between the observer and the dwarf, the two celestial coordinates return the projected radial distance $R$ of the star from the centre of the dwarf. Similarly, by assuming that all the stars are at the same distance from the observer, namely by neglecting the size of the dwarf along the line of sight, we can derive the three velocity components in the reference frame of the dwarf from the components of the proper motion and from the line-of-sight velocity component.

In our mock catalogues, for each star, we list the projected radial distance $R$ and the three velocity components $v_x$, $v_y$, and $v_z$ in the reference frame of the dwarf. We thus assume that the distance to the dwarf is known without uncertainty and that we have already transformed the observable quantities into the phase-space coordinates in the dwarf reference frame.

Table \ref{ParamsTable} lists the parameters we adopt to build our galaxy models. The DM parameters and 
the Plummer scale length are from Table 3 and Table 1, respectively, of \cite{Walker2009}. These parameters correspond to Draco galaxy. We adopt this dwarf as a working example of the possible targets of future {\it Theia}-like astrometric missions  \citep{Malbet2016,Theia2017,Malbet2019,Malbet2021}.

We create the mock catalogues of a dwarf whose DM density profile has either a core ($\gamma=0$) or a cusp ($\gamma=1$). We consider three different values of the velocity anisotropy parameter $\beta$; these values  are consistent with $\beta=0.25^{+0.47}_{-1.38}$ as measured  by \citet{massari2020} for Draco. 

\begin{figure*}
\begin{center}
\includegraphics[width=1.8\columnwidth]{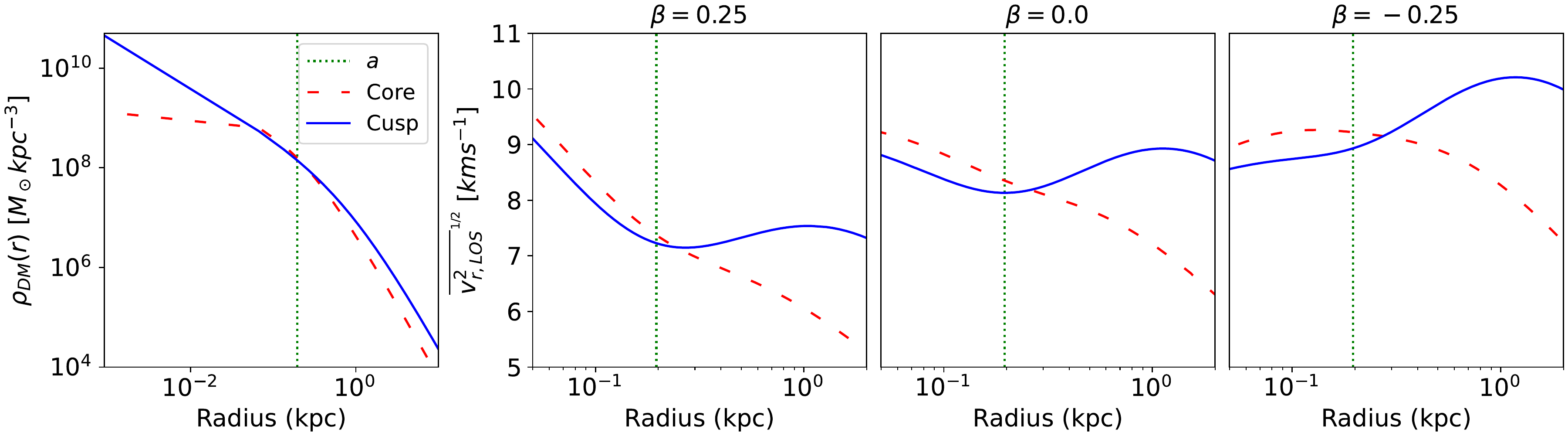}
\end{center}
\caption{{ Density profile (left panel) and velocity dispersion profiles projected along the line of sight (the three right panels) of our fiducial models. The solid blue and dashed red lines refer to the model with the central cusp and the central core respectively. The vertical green dotted lines show the location of the half-light radius $r_{1/2}=a$, where $a$ is the scale radius of the Plummer model of Eq.~(\ref{nup}). The value of the velocity anisotropy parameter $\beta$ is indicated on top of each of the three right panels.}}
\label{fig:ref3}
\end{figure*}

{ Figure \ref{fig:ref3} shows the density and velocity dispersion profiles of our models.  The density profile in the left panel shows that the two density profiles differ by 10\% at the half-light radius $r_{1/2}=a$. 
The profiles of the velocity dispersion projected along the line of sight differ, in the two models, by $\sim 2$\%, $\sim 2.6$\% and $\sim 3$\% within $r_{1/2}$ when $\beta=0.25, \, 0.0\,$ and  $-0.25$ respectively; the difference  substantially increases at larger radii. We do not show the profiles of the velocity dispersion of the proper motions which are qualitatively similar to the profiles of the velocity dispersion along the line of sight. The difference in the velocity dispersion profiles at large radii shows the relevance of the measure of the velocity field to distinguish between core and cusp models.  }

We consider five different sizes of the star samples: $N=100$, 1000, 2000, 4000, and 6000. The largest sample  {size,} $N=6000$, is set according to Z16, 
who found that this minimum number of stars is
required to distinguish between a core and a cusp if the line-of-sight velocities alone are available and the sample can be separated into two stellar populations according to the metallicity of the stars. 

Finally, we consider three different observational uncertainties $\sigma_v$ on the velocity measures, namely 0.0, 1.0, and 5.0 km~s$^{-1}$. 
For any star in our sample,  these uncertainties are associated to every Cartesian velocity component. This assumption is clearly unrealistic, and is made for the sake of simplicity. Indeed, (i) the uncertainties on the velocity components of a star depend on the star brightness, and (ii) different velocity components are measured with different techniques, either astrometric or spectroscopic, and thus suffer from different uncertainties. 

At the distance of Draco, $\sim 76$~kpc \citep[e.g.,][]{McConnachie2012}, the uncertainty $\sigma_v=1.0$~km~s$^{-1}$ implies an uncertainty $\sim 4\, \mu$as~yr$^{-1}$ on the proper motion of a star. This uncertainty is comparable to the precision of a few $\mu$as~yr$^{-1}$ expected for {\it Theia}-like astrometric missions \citep{Malbet2019,Malbet2021}.

\section{Data processing}\label{sec:dataprocessing}

\begin{table}
\caption{Priors of our Bayesian procedure adopted to recover the parameters of our dwarf model.  
$U$ indicates a uniform distribution in the range indicated in brackets.}
\centering
\begin{tabular}{cc|cc}
\hline
\multicolumn{2}{c}{ Parameter}       & \multicolumn{2}{c}{ Prior}  \\
\hline
$ \log_{10}\biggl(\frac{\rho_0}{ 10^{7} \rm{M}_{\odot}/\rm{kpc}^{3}}\biggr)$ & {DM central density}  & \multicolumn{2}{c}{$U({-3}, 5)$}    \\
{$ \log_{10}\biggl( \frac{r_{s}}{\rm{kpc}}\biggr)$}   & {DM scale radius}   &  \multicolumn{2}{c}{$U({-2}, 3)$} 
\\   
$\gamma$  & {central DM slope} & \multicolumn{2}{c}{${f{U(-3, 3)}}$}            \\
$\beta$   & {velocity anisotropy }      &   \multicolumn{2}{c}{$U(-10, 1)$}\\
\hline
\end{tabular}
\label{TablePriors}
\end{table}

The goal of our analysis is to determine the minimum size of the sample of stars and the minimum uncertainty on the proper motions that are required to properly recover the parameters of the DM distribution. We adopt a Bayesian approach. 
 
We use the Monte-Carlo-Markov-Chain (MCMC) software {\texttt emcee} \citep{Foreman2013}. 
The four parameters that completely describe our models are the DM central density, $\rho_0$, the DM scale length, $r_s$, the slope of the inner DM density profile, $\gamma$, and the velocity anisotropy, $\beta$. They define the four-dimensional vector $\mathbf{f}=(\rho_0,r_s,\gamma,\beta)$. Note that, with our simplified approach, we do not explore the determination of the slope parameters $\alpha$ and $\delta$ [Eq.~\eqref{DMdensityEqn}] that we unrealistically assume to be known.

 To speed up the chain convergence and reduce the burn-in phase, we first run 100 chains with randomly chosen starting points in the  parameter space. For each component of $\mathbf{f}$, we adopt a flat prior distribution within the ranges listed in Table \ref{TablePriors}. We run all the chains for 500 steps; we then identify the point in the four-dimensional parameter space that, over all the 100 chains, has the largest probability. 
We consider a four-dimensional cubic volume centered on this point.  In each dimension, the cube has { a} side $0.2$ times the value of the coordinate of the point in that dimension. 
Finally, we randomly choose $N_c=32$ points within this volume. We adopt these points as the starting points of the 32 new chains of our final analysis. We  verify that this initial procedure speeds up the  convergence of the 32 chains without affecting the estimation of the posterior distributions of the model parameters.  

 We guarantee the convergence of all the 32 chains at the same time as follows. For each $j$-th chain, at each step $N$, we consider the average value $\mu_\mathbf{f}^j$ of the values $\mathbf{f}_i^j$ estimated in all the previous steps, 
\begin{equation}
 \mu_\mathbf{f}^j  = 
 \frac{1}{N}\sum_{i=1}^{N} \mathbf{f}_i^j\, .
\end{equation}
We then estimate the autocorrelation function as a function of the lag $\tau$, 
\begin{equation}
\hat{C}_\mathbf{f}^j(\tau) = \frac{1}{N - \tau} \sum_{i=1}^{N-\tau} \bigl(\mathbf{f}_i^j - \mu_{\mathbf{f}}^j\bigr)\,\bigl(\mathbf{f}_{i+\tau}^j-\mu_{\mathbf{f}}^j\bigr)\,,
    \end{equation}
that,  in turn, provides the normalized autocorrelation function  averaged over the number of chains $N_c$,
\begin{equation}
 \hat{\rho}_\mathbf{f}(\tau) = \frac{1}{N_c}\sum_{j=1}^{N_c} \frac{\hat{C}_\mathbf{f}^j(\tau)}{\hat{C}_\mathbf{f}^j(0)}\,.
\end{equation}
We can now estimate the autocorrelation time for the vector $\mathbf{f}$,
\begin{equation}
\hat{\tau}_\mathbf{f} (M) = 1 + 2\,\sum_{\tau=1}^M \hat{\rho}_\mathbf{f}(\tau)\,,    
\end{equation}
where $M<N$. 

We check $\hat{\tau}_\mathbf{f} $ every 100 steps. The chains converge when two conditions are satisfied at the same time and for all  the components of $\mathbf{f}$: (1) all the chains are longer than 100 times  the  autocorrelation time, and (2) the autocorrelation time has changed by less than 1\%. 

The likelihood function is  
\begin{equation}\label{Prob}
\mathcal{L} = \prod^{n}_{i=1}  p({\boldsymbol v}_{i}| {\boldsymbol r}_{i}) \, ,
\end{equation}
where $i$ denotes the $i$-th star in the data set, and 
\begin{equation}\label{Likelihood2}
p({ v}| {\boldsymbol r}) = \frac{\exp \left\{ -\frac{1}{2}\left[ { v} - {\boldsymbol \mu({\boldsymbol r})} \right]^{T} \left[ { C({\boldsymbol r})} + { S({\boldsymbol r})} \right]^{-1}\left[ { v} - {\boldsymbol \mu({\boldsymbol r})} \right]\right\} }{\sqrt{\left( 2\pi \right)^{n}\left| { C({\boldsymbol r})} + { S({\boldsymbol r})} \right|}}
\end{equation} 
is the convolution between the Gaussian distribution with covariance matrix $\mathbf {S(r)}$, representing the instrumental errors, and the Gaussian distribution with covariance matrix $\mathbf {C(r)}$ representing the probability distribution of the velocity components given in Eq.~\eqref{Likelihood1}.

If the measurements are uncorrelated, the off-diagonal elements of $\mathbf {S ( r)}$ are {equal to} zero and the diagonal components of $\mathbf {S ( r)}$ are the squared errors of the error measurements: $\{ \sigma_{{\rm{v}}_{x}}^{2},\sigma_{{\rm{v}}_{y}}^{2},\sigma_{{\rm{v}}_{z}}^{2}\}$. We assume no error on the positions.

As anticipated in Sect.~\ref{sec:astro_mocks}, we consider mock catalogues with velocity uncertainty $\sigma_v=0.0$, $1.0$, or $5.0$~km~s$^{-1}$ on each of the three Cartesian velocity components: $\sigma_{{\rm{v}}_{x}}=\sigma_{{\rm{v}}_{y}}=\sigma_{{\rm{v}}_{z}}\equiv \sigma_v$.  We do not model any dependence of the velocity uncertainty on the star magnitude.
For reference, Z16 perturbed the stars in their sample with an error of 3 km~s$^{-1}$.

\section{Results}\label{sec:results_sec}

Here, we show the results of the MCMC analysis of our mock catalogues. As a test case, in Sect. \ref{sec:CIDG} we show the results of a dwarf galaxy with a core, $\gamma=0$, and isotropic velocity field, $\beta=0$: this analysis assesses the constraining power of the proper motions of the stars. 
We then explore how well our MCMC approach can estimate the model parameters for galaxies with either a cusp or a core and with three different values of the velocity anisotropy parameter $\beta$. In Sect.~\ref{sec:CCP}, we explore how 
the measurement of the proper motion is relevant to solve the CCP. 

\subsection{Estimation of the parameters of the DM distribution and of the velocity field}\label{sec:CIDG}

\begin{figure*}
\begin{center}
\begin{tabular}{c}
\includegraphics[width=1.99\columnwidth]{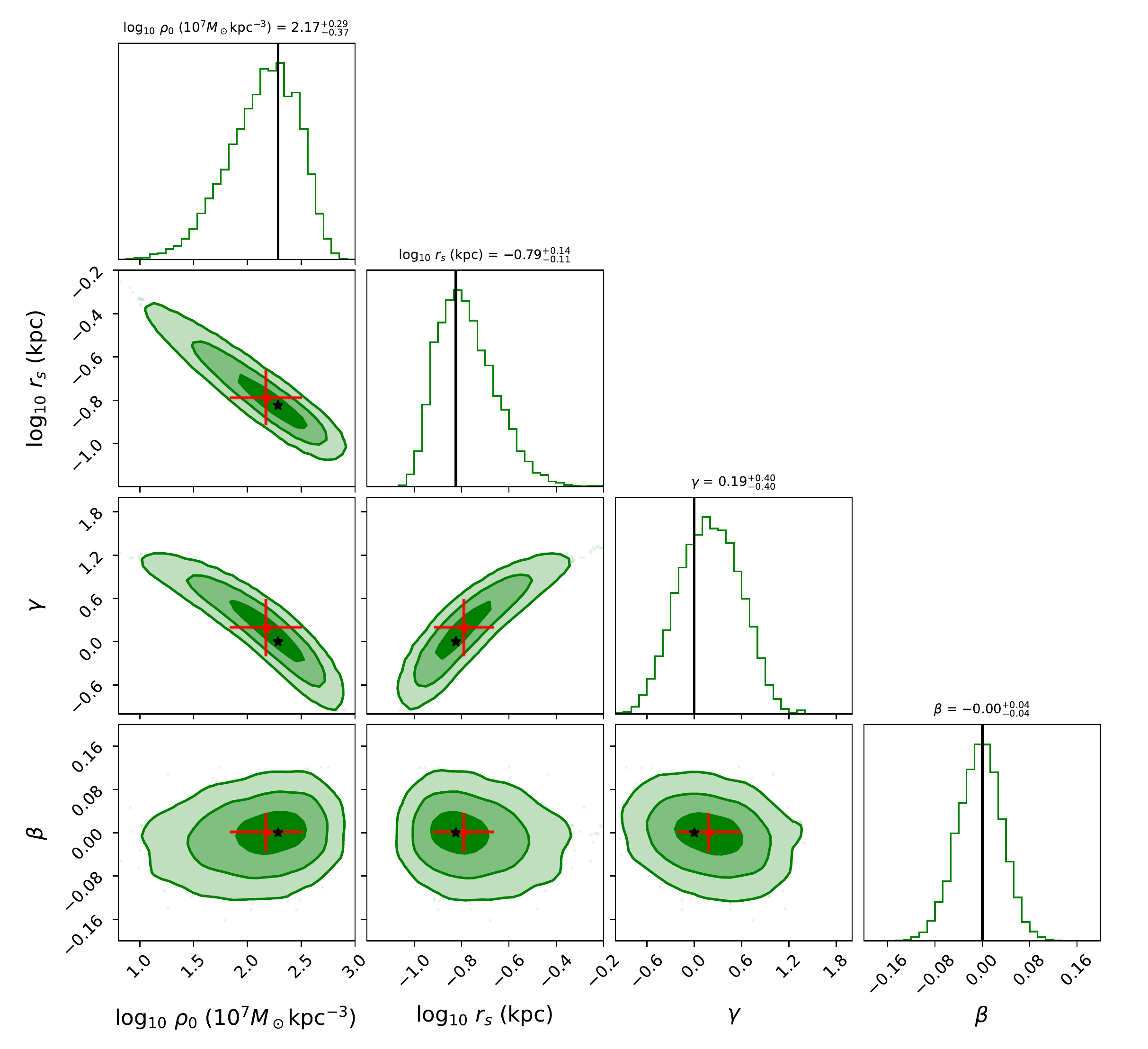}
\end{tabular}
\end{center}
\caption{MCMC posterior distributions of the parameters of  the  model with a core, $\gamma=0$, and an isotropic velocity field, $\beta=0$, estimated from a mock sample with $N=2000$ stars and uncertainty on each velocity component of the stars  $\sigma_v=0$~km~s$^{-1}$. 
The shaded areas with increasing darkness show the 99\%, 95\%, and 68\% confidence regions of the posterior distributions, respectively. The black solid stars show the position of the input parameters used to create the mock catalogue. The input parameters are also indicated by the vertical lines in the top panel of each column. The red solid circles  with the  error bars are the medians of the posterior distributions with their 68\% confidence intervals; they are adopted as the estimated values of the parameters and their uncertainty. These values are reported at the top of each column. 
}
\label{fig:2}
\end{figure*}

\begin{table*}
\caption{Parameters of the model with a core, $\gamma=0$, and an isotropic velocity field, $\beta=0$, estimated with our MCMC analysis. The estimates are the medians of the posterior distributions. The uncertainties correspond to the 68\% confidence intervals of the posterior distributions. $N$ is the number of stars in the mock catalogue; $\sigma_v$ is the uncertainty on the measure of each velocity component of the stars in the mock catalogues. }\label{tab:table_mcmc_fit}
  \begin{tabular}{lcc|ccccc}
    \toprule
    {Parameter} & {Input value} & ${ \sigma}_v$ (km/s) & {$N=100$} & {$N=1000$} & {$N=2000$} & {$N=4000$} & {$N=6000$}\\
    \midrule
    \multirow{4}{*}{$\log_{10}\biggl(\frac{\rho_{0}}{ 10^7 \rm{M}_{\odot}/\rm{kpc}^{3}}\biggr)$} & \multirow{4}{*}{2.284} & 0.0 & $0.66^{+1.07}_{-0.73}$ & $1.22^{+0.60}_{-0.54}$ & $2.17^{+0.29}_{-0.37}$ & $1.97^{+0.29}_{-0.29}$ & $2.15^{+0.21}_{-0.23}$  \\[0.1cm]
    &  & 1.0 & $0.64^{+1.08}_{-0.74}$ & $1.22^{+0.62}_{-0.60}$ & $2.18^{+0.29}_{-0.38}$ & $1.97^{+0.29}_{-0.30}$ & $2.15^{+0.22}_{-0.23}$  \\[0.1cm]
    &  & 5.0 & $0.99^{+1.20}_{-1.22}$ & $1.49^{+0.78}_{-1.01}$ & $2.25^{+0.29}_{-0.43}$ & $2.10^{+0.34}_{-0.44}$  & $2.17^{+0.29}_{-0.36}$ \\[0.1cm]
     \midrule
        \multirow{4}{*}{$\log_{10}\biggl(\frac{r_s}{ \rm{kpc}}\biggr)$} & \multirow{4}{*}{-0.824} & 0.0 & $-0.21^{+0.31}_{-0.43}$ & $-0.42^{+0.22}_{-0.24}$ & $-0.79^{+0.14}_{-0.11}$ & $-0.70^{+0.11}_{-0.11}$ & $-0.78^{+0.09}_{-0.08}$  \\[0.1cm]
    &  & 1.0 & $-0.20^{+0.31}_{-0.43}$ & $-0.42^{+0.25}_{-0.24}$ & $-0.80^{+0.14}_{-0.11}$ & $-0.70^{+0.11}_{-0.11}$ & $-0.78^{+0.09}_{-0.08}$   \\[0.1cm]
    &  & 5.0 & $-0.55^{+0.49}_{-0.46}$ & $-0.66^{+0.41}_{-0.30}$  & $-0.94^{+0.16}_{-0.11}$ & $-0.86^{+0.17}_{-0.12}$ & $-0.90^{+0.13}_{-0.10}$  \\[0.1cm]
     \midrule
         \multirow{4}{*}{$\gamma$} & \multirow{4}{*}{0.0} & 0.0 & $1.29^{+0.28}_{-0.65}$ & $0.99^{+0.28}_{-0.46}$ & $0.19^{+0.40}_{-0.40}$ & $0.33^{+0.29}_{-0.36}$ &  $0.20^{+0.26}_{-0.28}$  \\[0.1cm]
    &  & 1.0 & $1.30^{+0.28}_{-0.63}$ & $0.99^{+0.30}_{-0.48}$ & $0.19^{+0.41}_{-0.40}$ & $0.33^{+0.29}_{-0.36}$ & $0.20^{+0.26}_{-0.29}$   \\[0.1cm]
    &  & 5.0 & $1.47^{+0.46}_{-0.79}$ & $0.89^{+0.53}_{-0.77}$& $0.16^{+0.47}_{-0.43}$  & $0.17^{+0.48}_{-0.48}$ & $0.20^{+0.41}_{-0.40}$  \\[0.1cm]
     \midrule
        \multirow{4}{*}{$\beta$} & \multirow{4}{*}{0.0} & 0.0 & $-0.37^{+0.22}_{-0.25}$ & $-0.09^{+0.06}_{-0.06}$ & $0.00^{+0.04}_{-0.04}$ & $-0.01^{+0.03}_{-0.02}$ & $0.00^{+0.02}_{-0.02}$   \\[0.1cm]
    &  & 1.0 & $-0.38^{+0.22}_{-0.26}$ & $-0.09^{+0.06}_{-0.06}$ & $0.00^{+0.04}_{-0.04}$ & $-0.01^{+0.03}_{-0.03}$ & $0.00^{+0.02}_{-0.02}$   \\[0.1cm]
    &  & 5.0 & $-0.88^{+0.51}_{-0.61}$ & $-0.15^{+0.09}_{-0.10}$ & $-0.02^{+0.06}_{-0.06}$ & $-0.01^{+0.04}_{-0.04}$  & $0.00^{+0.03}_{-0.03}$  \\[0.1cm]
    \bottomrule
  \end{tabular}
\end{table*}

In our MCMC analysis, we run 32 chains with random starting points in { the} parameter space as illustrated in the previous section. Examples of the post burn-in posterior density distributions of the parameters are shown in Fig. \ref{fig:2}. This figure shows the case of a mock catalogue with 2000 stars and no errors on the velocity measures.
Table \ref{tab:table_mcmc_fit} lists the results of the MCMC analyses of our full set of mock catalogues for a dwarf with a core, $\gamma=0$, and an isotropic velocity field, $\beta=0$. We adopt  the medians of the posterior distributions as the best estimates of the parameters; the 15.9 and 84.1 percentiles of the posterior distributions yield the $1\sigma$  
uncertainties of these estimates. 

Figure~\ref{fig:2} shows that the posterior distributions of the individual parameters are approximately Gaussian, although some of the distributions show substantial skewness and extended tails; the parameters of the density profile, $\rho_0$, $r_s$, and $\gamma$, are degenerate. The degeneracy between $\rho_0$ and $r_s$ is in agreement with the results of Z16.
Nevertheless, for $N\ge 2000$ stars, the MCMC analysis returns the estimates of these parameters within $1\sigma$ of the input value (Table \ref{tab:table_mcmc_fit}).   

\begin{figure}
\begin{center}
\includegraphics[width=1.1\columnwidth]{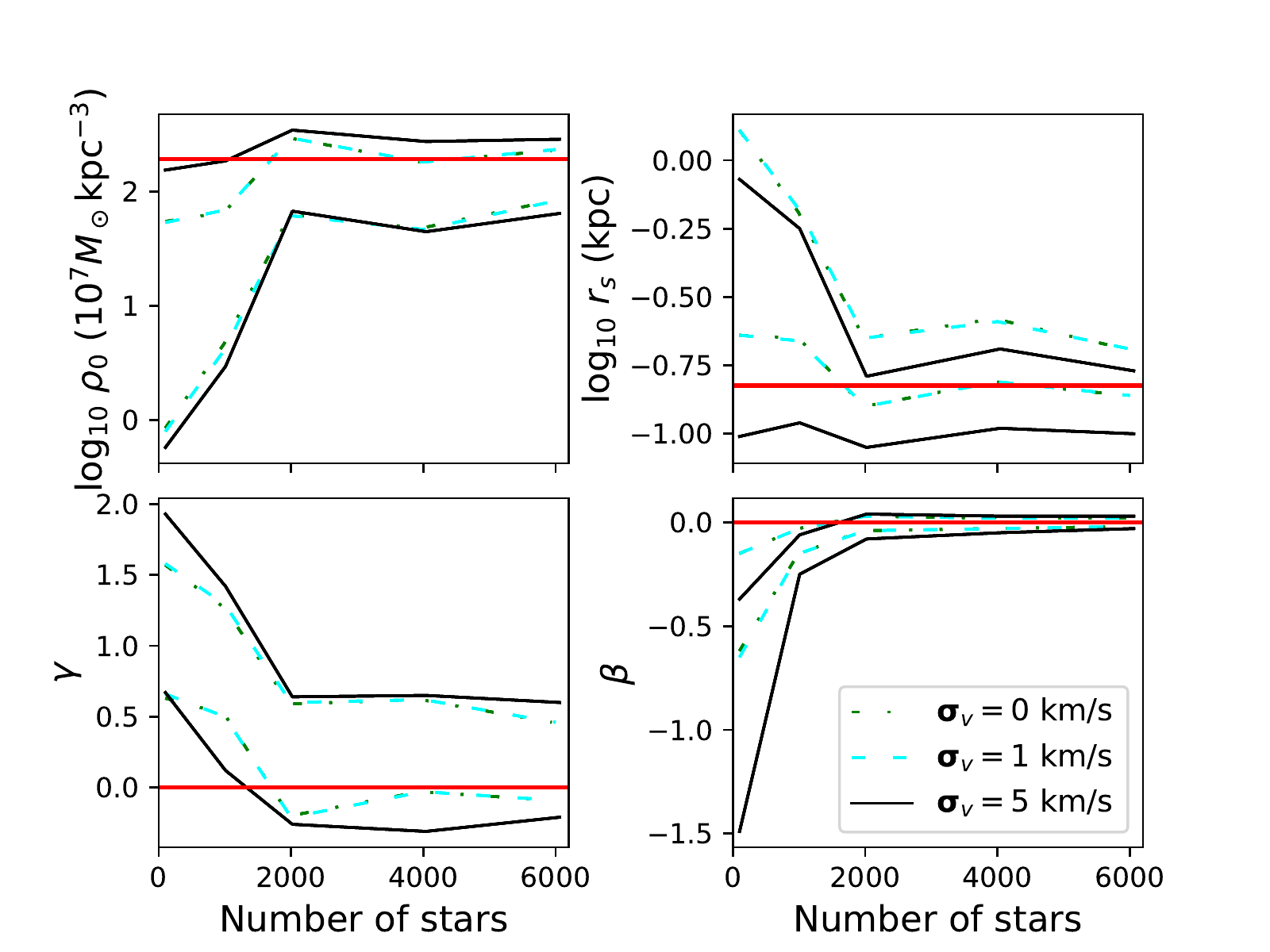}
\end{center}
\caption{The $1\sigma$ confidence interval (upper and lower curves with the same style) of the parameter estimates as a function of the number of stars of the catalogue for a model with a core, $\gamma=0$, and an isotropic velocity field, $\beta=0$. Different line styles are for different  velocity uncertainty $\sigma_v$, as listed in the inset. The red horizontal lines show the input parameters used to create the mock catalogues. The confidence intervals for $\sigma_v=0$ km s$^{-1}$ and $\sigma_v=1$ km s$^{-1}$ are indistinguishable. 
}
\label{fig:3}
\end{figure}

Figure \ref{fig:3} shows how the $1\sigma$ confidence interval of the parameter estimates depends on the size $N$ of the star catalogue and on the uncertainty $\sigma_v$  of the velocity measures. Clearly, when $N\le 1000$, the MCMC analysis is unable to recover the input parameters. On the contrary, the estimated parameters  converge to the input values when $N\ge 2000$. The convergence is more evident for the velocity anisotropy parameter $\beta$.  

By measuring the proper motions of 45 stars in Draco, with the combination of {\it Hubble Space Telescope} observations  and  the second {\it Gaia} Data  Release, \citet{massari2020} estimate the velocity anisotropy parameter $\beta=0.25^{+0.47}_{-1.38}$. These uncertainties agree with the expected errors on $\beta$ for our mock catalogue with $N=100$ stars. Our analysis shows that reducing the uncertainty by a factor of ten requires a number of stars at least 20 times larger.

Table \ref{tab:table_mcmc_fit} shows that, when $N\ge 2000$,  the velocity uncertainty $\sigma_v$ within the range we explore has a moderate impact on the uncertainty of the parameter estimates. For these large sample sizes, the estimated values are unbiased and within $\sim 10$\% of the input values; in addition, the $1\sigma$  uncertainties are generally smaller than $\sim 20$\%.  

\begin{table}
\caption{Same as Table \ref{tab:table_mcmc_fit} for models with either a core, $\gamma=0$, or a cusp, $\gamma=1$,  and three different values of the velocity anisotropy parameter $\beta$. All the mock catalogues in this table contain $N=2000$ stars and have velocity uncertainty $\sigma_v=0$ km s$^{-1}$.  The input values of $\log_{10}(\rho_0/ 10^7 \rm{M}_\odot \rm{kpc}^{-3})$ are $[2.28, 0.73]$ for the models with a core and with a cusp, respectively. The input values of  $\log_{10}(r_s/ \rm{kpc})$ are 
$[-0.82,-0.10]$, respectively. }\label{tab:table_mcmc_fit_core_cuspy}
  \begin{tabular}{lc|cc}
    \toprule
    { Parameters} & { Input}  $\beta$ & { Core ($\gamma=0$)} & { Cusp ($\gamma=1$)} \\
    \midrule
    \multirow{4}{*}{$\log_{10}\biggl(\frac{\rho_{0}}{ 10^{7} \rm{M}_{\odot}/\rm{kpc}^{3}}\biggr)$} &  -0.25 & $2.04^{+0.30}_{-0.34}$ & $0.71^{+0.21}_{-0.25}$ \\[0.1cm]
    &  0.0 & $2.17^{+0.29}_{-0.37}$  & $0.75^{+0.21}_{-0.24}$   \\[0.1cm]
    &  0.25 & $2.51^{+0.18}_{-0.21}$  & $0.43^{+0.27}_{-0.28}$   \\[0.1cm]
     \midrule
        \multirow{4}{*}{$\log_{10}\biggl(\frac{r_{s}}{ \rm{kpc}}\biggr)$} & -0.25 & $-0.76^{+0.13}_{-0.11}$ & $-0.11^{+0.11}_{-0.09}$   \\[0.1cm]
    &   0.0 & $-0.79^{+0.14}_{-0.11}$ & $-0.10^{+0.11}_{-0.09}$  \\[0.1cm]
    &   0.25 & $-0.90^{+0.08}_{-0.07}$ & $0.03^{+0.13}_{-0.12}$  \\[0.1cm]
     \midrule
\multirow{4}{*}{$\gamma$} &  -0.25 & $0.40^{+0.32}_{-0.37}$ & $1.08^{+0.10}_{-0.11}$  \\[0.1cm]
    &   0.0 & $0.19\pm0.40$ & $0.97\pm0.12$   \\[0.1cm]
    &  0.25 & $-0.41^{+0.34}_{-0.32}$ & $1.16^{+0.11}_{-0.12}$  \\[0.1cm]
         \midrule
             \multirow{4}{*}{$\beta$} &  -0.25 & $-0.35\pm0.06$ & $-032\pm0.05$  \\[0.1cm]
    &   0.0 & $0.0\pm0.04$ & $0.06\pm0.04$   \\[0.1cm]
    &  0.25 & $0.25\pm0.03$ & $0.21\pm0.03$  \\[0.1cm]
    \bottomrule
  \end{tabular}
\end{table}

\begin{figure}
\begin{center}
\includegraphics[width=1.1\columnwidth]{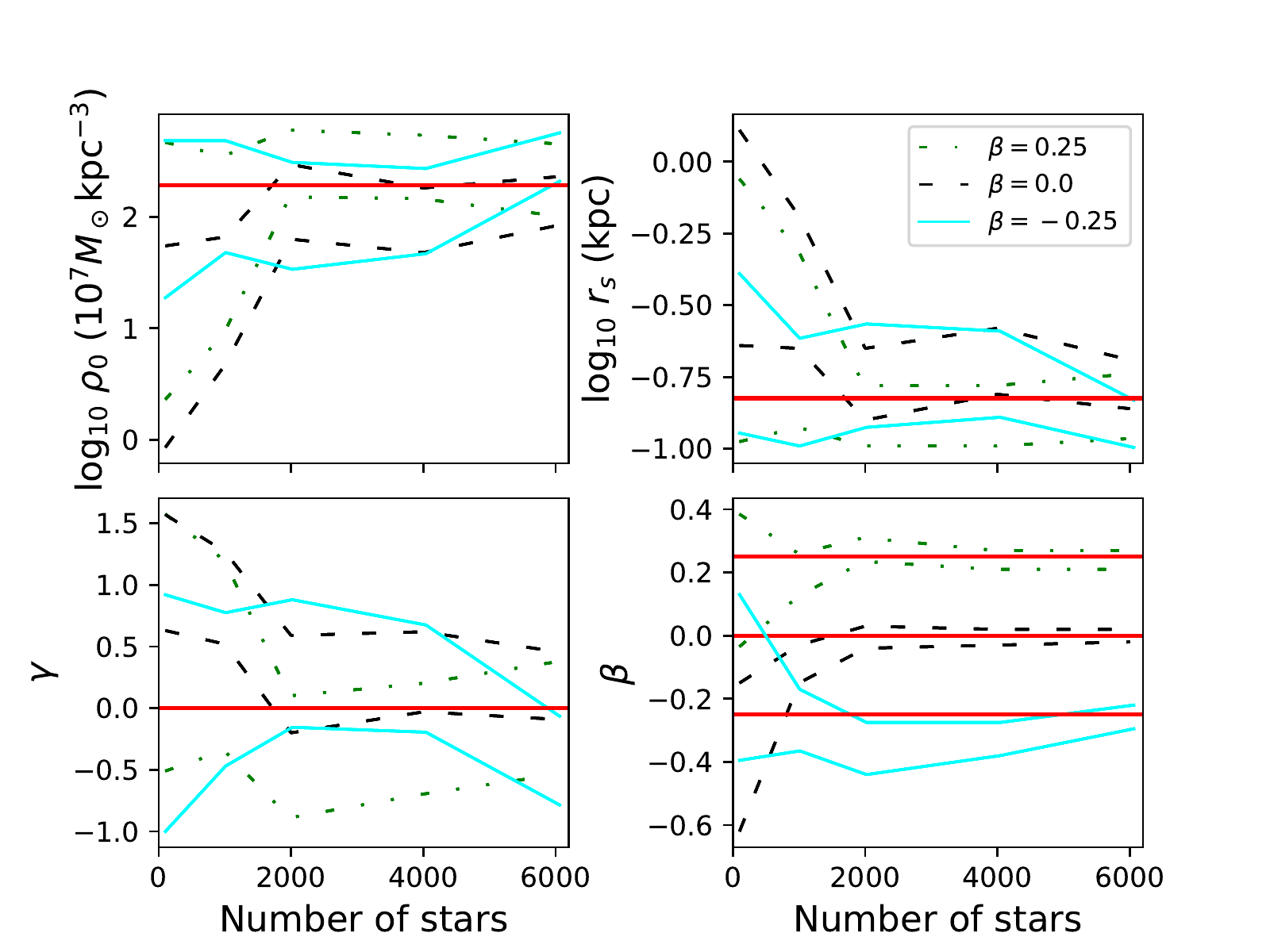}
\end{center}\caption{The $1\sigma$ confidence interval (upper and lower curves with the same style) of the parameter estimates as a function of the number of stars of the catalogue for models with a core, $\gamma=0$, and three different values of the velocity anisotropy parameter $\beta$, as listed in the inset. The red horizontal lines show the input parameters used to create the mock catalogues. In these mock catalogues, the  uncertainty on each velocity component of the individual stars is $\sigma_v=0$ km s$^{-1}$. 
}
\label{fig:4}
\end{figure}

\begin{figure}
\begin{center}
\includegraphics[width=1.1\columnwidth]{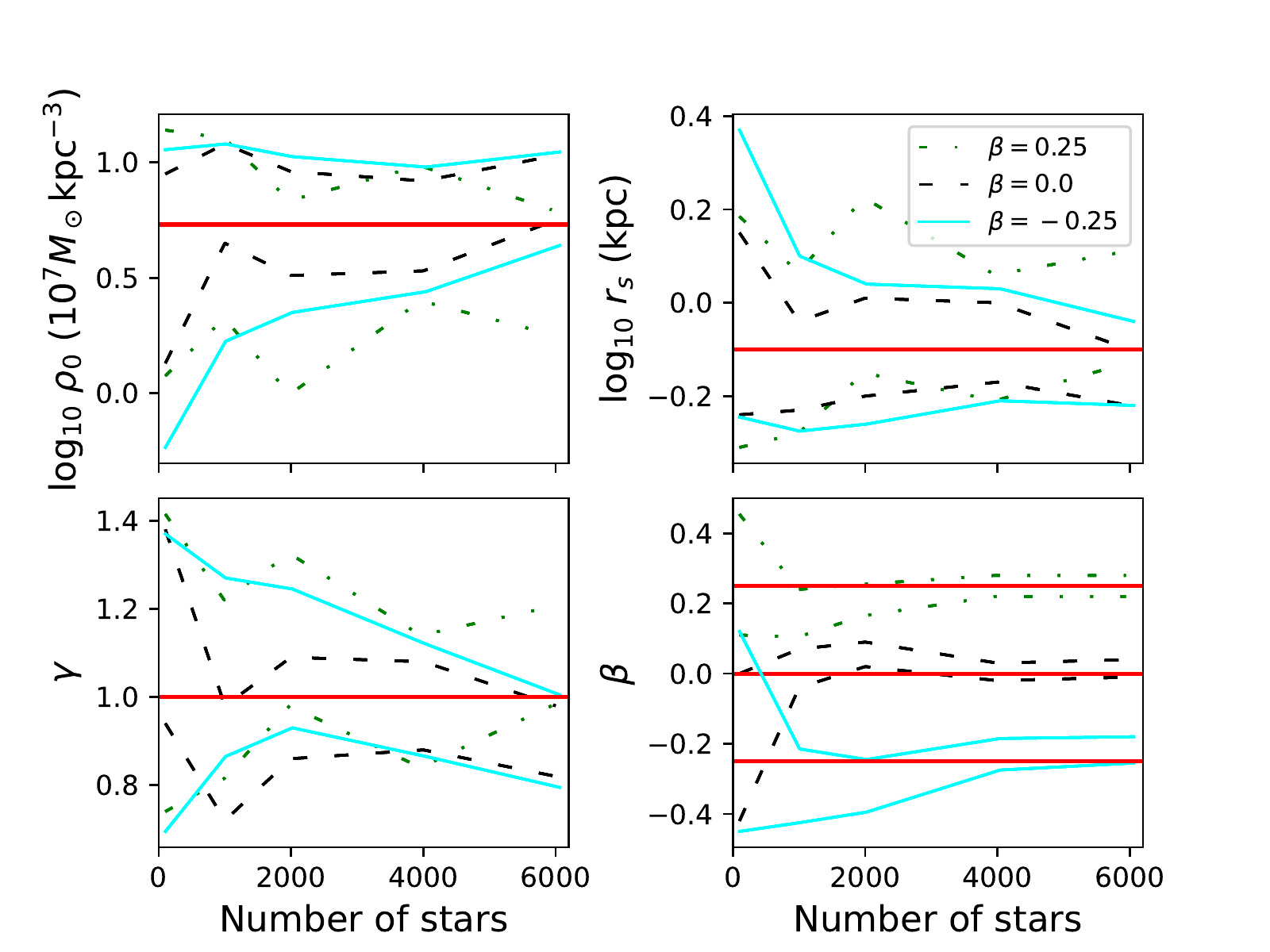}
\end{center}
\caption{Same as Fig. \ref{fig:4} for models with a cusp,  $\gamma=1$.
}
\label{fig:5}
\end{figure}

Figures \ref{fig:4} and \ref{fig:5} show the results obtained when we apply our methodology to our mock catalogues of a dwarf model with three different values of $\beta$ and with either a cusp or a core in the density profile. In addition, the results shown in these figures  assume $\sigma_v=0$ km~s$^{-1}$. According to Table  \ref{tab:table_mcmc_fit}, larger uncertainties on the velocities decrease the precision of the estimated parameters but leave  the estimates unbiased. 
The size of the star catalogues appears more relevant than the velocity uncertainty $\sigma_v$. 

Figures \ref{fig:4} and \ref{fig:5} show that, when the catalogue contains at least $N=2000$ stars, the input parameters are satisfactorily recovered for any model. Table \ref{tab:table_mcmc_fit_core_cuspy} lists the estimates of the parameters for the catalogues with $N=2000$ stars.

Table \ref{tab:table_mcmc_fit_core_cuspy} shows that samples of 2000 stars can return the estimate of the central density $\rho_0$ with a relative error $\sim 15\%$ if the density profile has a core, and a relative error $\sim 30-60\%$ if the density profile has a cusp. These uncertainties are consistent with those estimated by \citet{Read2018}, who combined $\sim 2500$ stars with photometric information and $\sim 500$ stars with measured line-of-sight velocity components to measure the DM density within $150$~pc of the centre of Draco. Our analysis suggests that these uncertainties can be reduced by a factor $\sim 1.5-2$, with a sample of   6000 stars.

\subsection{Distinguishing between a cusp and a core}\label{sec:CCP}

\begin{figure*}
\centering
\includegraphics[width=\columnwidth]{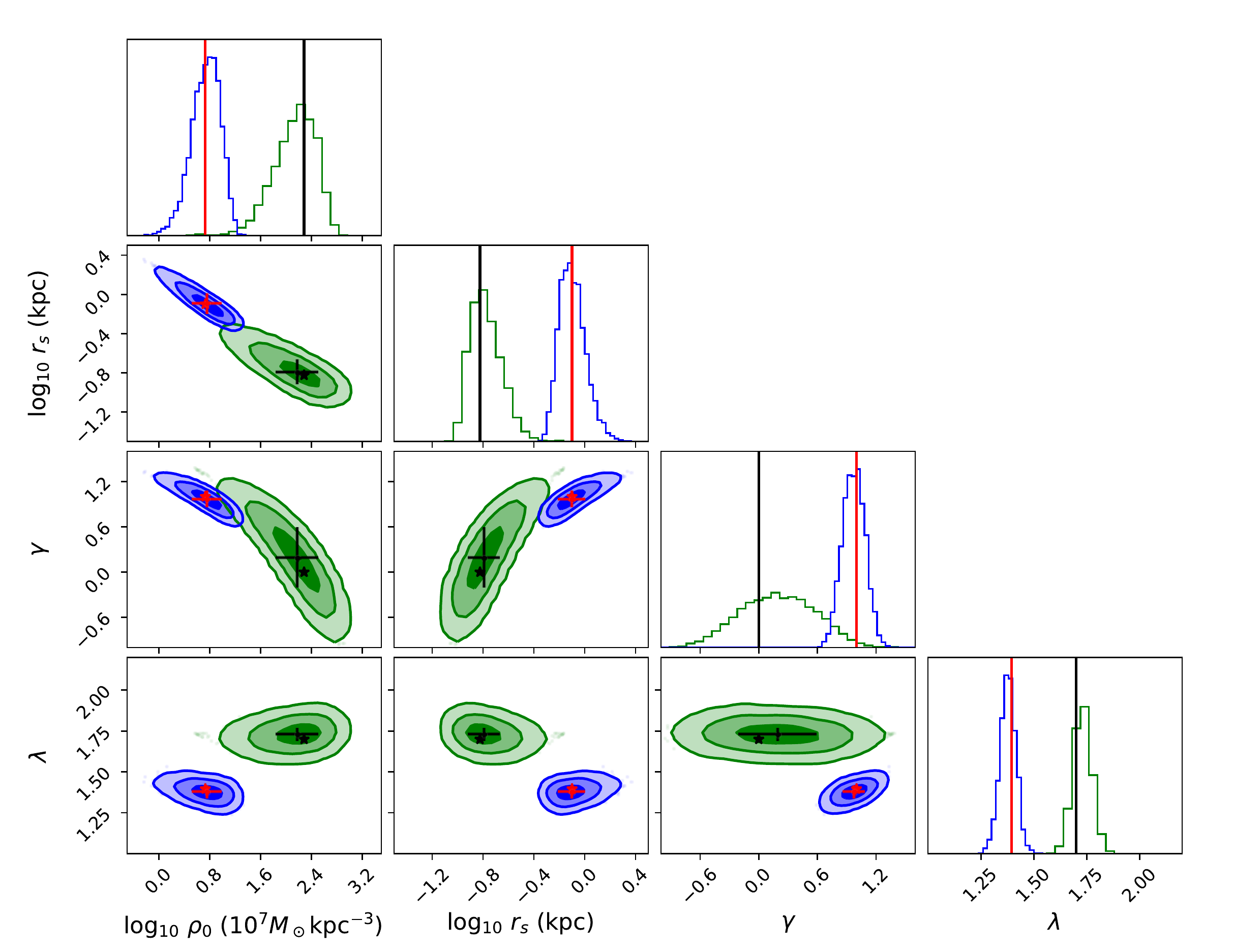}
\includegraphics[width=\columnwidth]{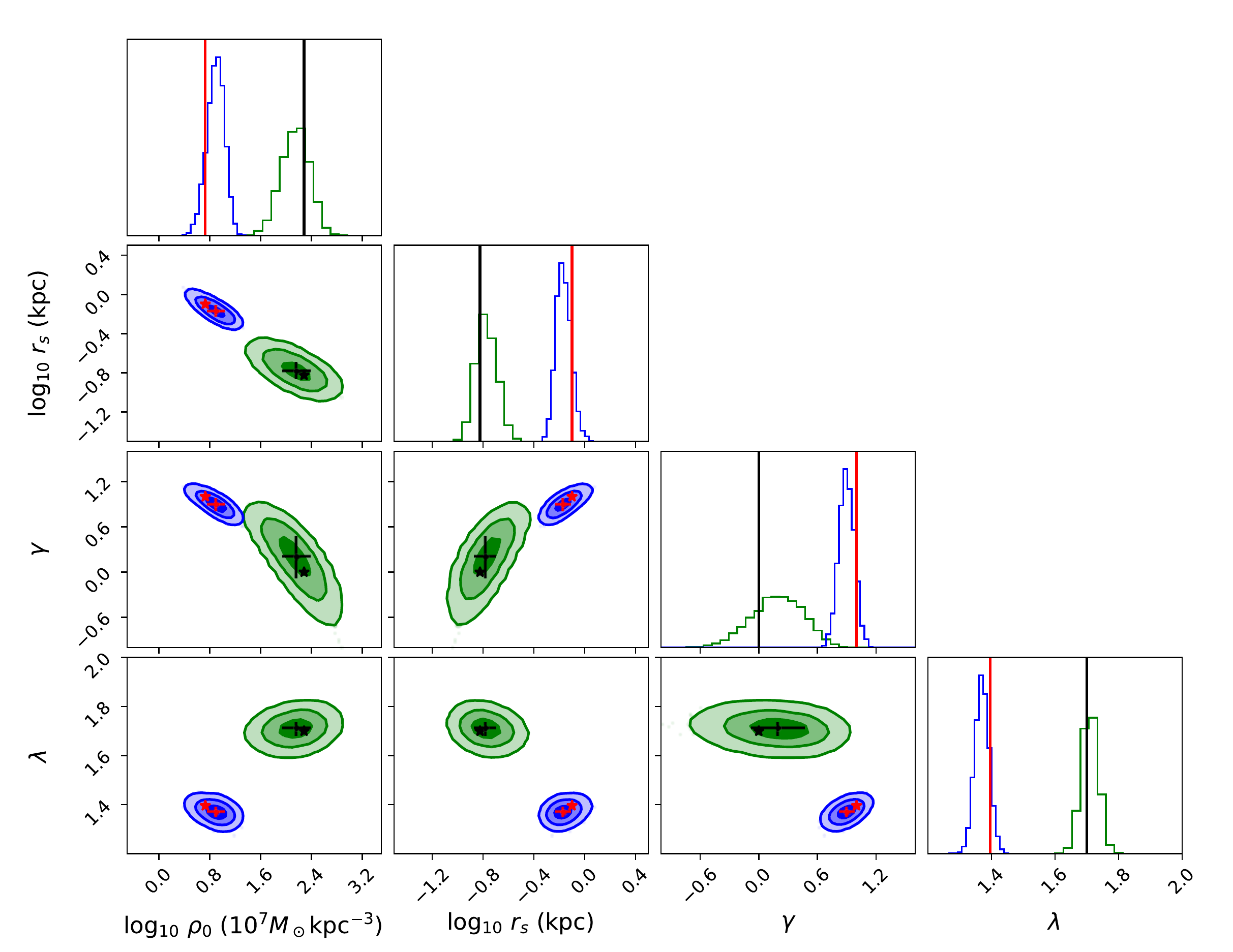}
\caption{Posterior distributions of the estimated parameters of the DM density profiles of a dwarf model with a cusp (blue shaded areas) or a core (green shaded areas) and velocity anisotropy parameter $\beta=0$. The left (right) panels show the posterior distributions derived from mock catalogues with $N=6000$ {{($\mathbf N=2000$)}} stars. We assume the uncertainty $\sigma_v=0$ km~s$^{-1}$ on the stellar velocity components.
The red and black solid stars and the red and black vertical lines indicate the input parameters of the models. The medians and the 68\% confidence intervals of the posterior distributions are shown by the red and black dots with error bars and reported in Table \ref{tab:table_ccp_results}.  In this table, the MCMC analysis based on the mock catalogues with $N=6000$ and {${N=2000}$} stars are  labeled 6000-3D and { 2000-3D}, respectively. } 
\label{ccp_corner}
\end{figure*}

We now show how the measures of the  proper motions of the stars can in principle distinguish between a cusp and a core independently of the specific value of the velocity anisotropy parameter. 
We consider two identical models except for the value of $\gamma$. We set the  velocity anisotropy parameter $\beta=0$, and the velocity uncertainties $\sigma_v=0$ km~s$^{-1}$. 
The green shaded areas in the left panels of Fig.~\ref{ccp_corner} show the posterior distributions of the recovered parameters of the DM density profile with a core, $\gamma=0$, derived with our MCMC analyses of a mock catalogue with $N=6000$ stars. The blue shaded areas show the results for the same case but with  a cusp, $\gamma=1$. 

For catalogues of this size, our MCMC analysis returns the estimates $\gamma=0.20\pm^{+0.26}_{-0.28}$ and $\gamma=0.90\pm0.08$ for the model with a core and with a cusp, respectively. In other words, this result indicates that our MCMC analysis { would distinguish between a cusp and a core:} 
if the dwarf has a cusp, a core can be ruled out at largely more than $3\sigma$. 
{ We can actually do better: according to the suggestion of \citet{Strigari2007} (see also \citealt{Guerra2021}), the log-slope $\lambda(r)=-{\mathrm d}\ln\rho/{\mathrm d}\ln r$, estimated at the half-light radius $r_{1/2}$, can distinguish the two density profiles more efficiently than the measure of $\gamma$ alone. For our density profile in Eq.~\eqref{DMdensityEqn}, $\lambda(r)=\gamma-(\gamma-\delta)(r/r_s)^\alpha/[1+(r/r_s)^\alpha]$, and for the Plummer stellar density distribution of Eq.~\eqref{nup}, $r_{1/2}=a$. In our fiducial models, we thus find 
$\lambda(a)=1.70$ and $1.40$ for the profiles with a core and a cusp, respectively. The last row of the panels of Fig. \ref{ccp_corner}  shows the posterior distributions of $\lambda$. With a sample of $6000$ stars, the two models can now be unambiguously distinguished at $16 \sigma$. } 

An anisotropic velocity field 
{ increases the separation between} the two density models. With a mock catalogue of 6000 stars with $\beta=-0.25$, we find $\gamma=-0.44^{+0.25}_{-0.23}$ and $\gamma=0.90\pm0.07$ for the model with a core and with a cusp, respectively. With $\beta=0.25$, we get $\gamma=-0.08\pm^{+0.30}_{-0.31}$ and $\gamma=1.10\pm0.08$, respectively.  {  For both values of $\beta$, we would thus distinguish the two models at more than $3\sigma$. As for $\lambda$, with $\beta=-0.25$, we find $\lambda=1.69^{+0.02}_{-0.03}$ and $\lambda=1.34\pm0.02$ for the model with a core and with a cusp, respectively. With $\beta=0.25$, we get $\lambda=1.73^{+0.02}_{-0.03}$ and $\lambda=1.42^{+0.02}_{-0.03}$, respectively.  For both values of $\beta$, we can now distinguish the two models at more than $10 \sigma$.}

The right panels of Fig. \ref{ccp_corner} show the MCMC analysis of the same two models but { with fewer stars, $N=2000$.
The estimates of the central slope of the density profiles are now 
$\gamma=0.19\pm0.40$ and $\gamma=0.97\pm 0.11$ for the model with a core and with a cusp, respectively.} With this smaller sample of stars, { limiting the analysis to the values of $\gamma$ might thus distinguish the two models at $\sim 2\sigma$. On the contrary, the measure of the log-slope, $\lambda=1.73\pm0.04$ and $1.38\pm0.04$ for the model with a core and a cusp respectively, distinguishes the two models at more than $8\sigma$.} 
{ The separation between the posterior distributions of $\lambda$ shown in the right panel of Fig. \ref{ccp_corner} might suggest that a  sample size even smaller than $N=2000$ would distinguish a core from a cusp. However, samples with smaller $N$ return strongly biased values of the model parameters (see Sect. \ref{sec:CIDG}). 
Therefore, samples with $N<2000$ stars are not recommended.}

We conclude that our approach will be able to  unambiguously { distinguish between cusp and core models} for a Draco-like galaxy if we measure the proper motions of at least { 2000}  
stars, a sample size that is within the capability of future {\it Theia}-like astrometric missions  \citep{Theia2017,Malbet2021} { or ground-based 30 meter class telescopes \citep{Skidmore2015}. }

{ Our limit $N=2000$ for the size of the star sample is larger than the limit derived by the analysis of \citet{Guerra2021} who find that $N=1000$ is sufficient to distinguish between a core and a cusp at the $2\sigma$ confidence level. However, the result of \citet{Guerra2021} depends on the details of the density profile and such a distinction is thus not always guaranteed. \citet{Guerra2021} use a Fisher matrix formalism validated by the DYNESTY code \citep{Speagle2020}, which is based on the dynamic nested sampling method of \citet{Higson2019}. We use a traditional MCMC approach and we indeed confirm that a core and a cusp can in principle be separated with $N<2000$ stars. However, as mentioned above, with these poorer samples the parameters of the density profile are strongly biased and we are unable to recover the correct mass density distribution and velocity anisotropy parameter.  }

\begin{figure}
\begin{center}
\includegraphics[width=\columnwidth]{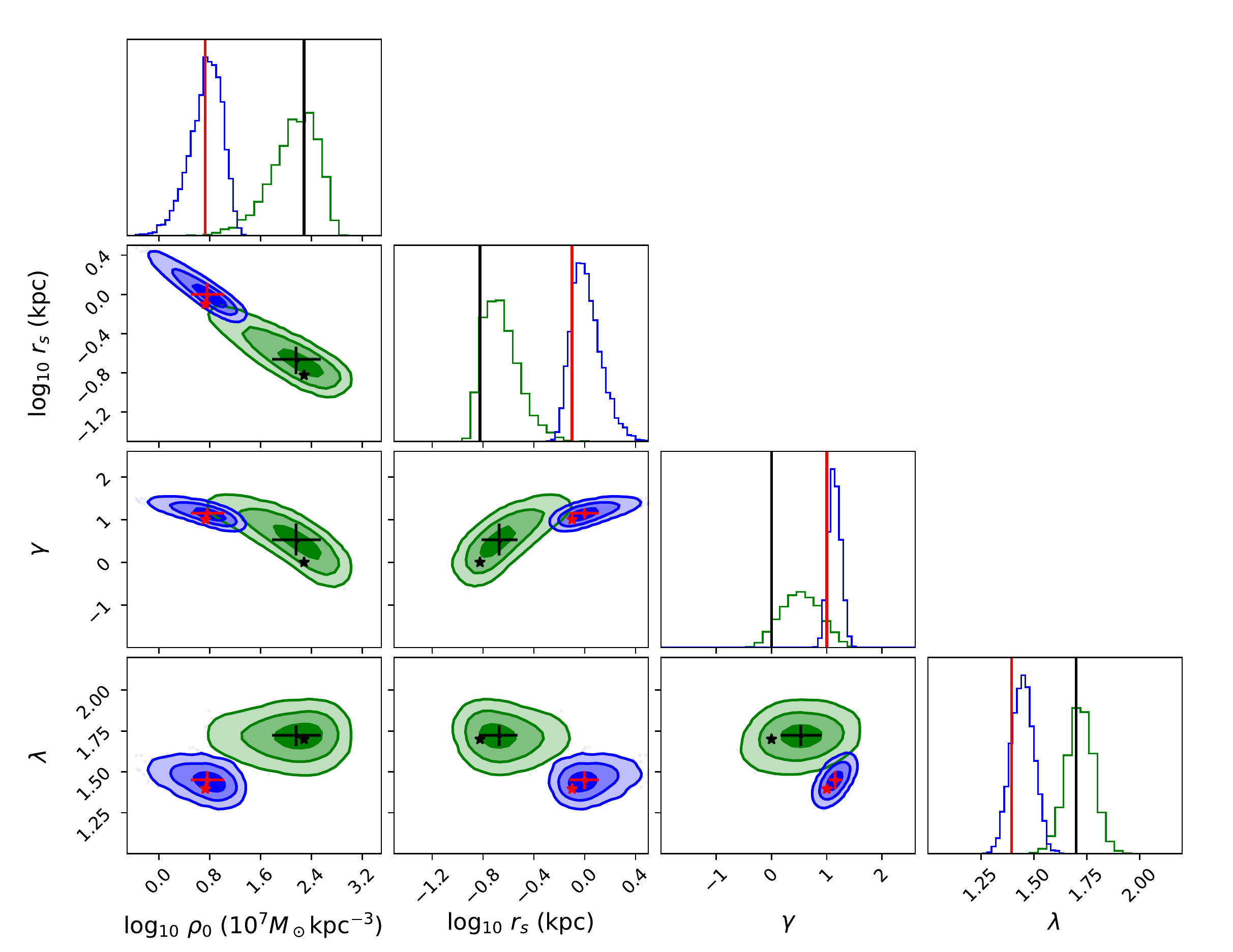}
\end{center}
\caption{Same as Fig. \ref{ccp_corner} for a mock catalogue with $N=6000$ stars where the proper motions of the stars are unavailable and only the line-of-sight components of the velocities are known. Input and estimated parameters for this model labeled 6000-1D are listed in Table \ref{tab:table_ccp_results}. 
} 
\label{ccp_corner_1D}
\end{figure}

The relevance of the measures of the proper motions is shown in Fig. \ref{ccp_corner_1D}. We show the results of the MCMC analysis of the mock catalogues with $N=6000$ stars of the two models with a cusp or with a core where only the line-of-sight velocities of the stars are known. Compared to the case with $N=6000$, where all the three velocity components are known (left panels of Fig.  \ref{ccp_corner}), the posterior distributions for the { parameters $\gamma$ and $\lambda$} now degrade and become { roughly} comparable to the distributions for  the catalogues with 
${ {N=2000}}$ stars  where all the three velocity components are known (right panels of Fig. \ref{ccp_corner}). { Nevertheless, the measure of $\lambda$ can still distinguish the two models at more than $4\sigma$. Samples with $N< 6000$ stars with unknown proper motions are thus unable to unambiguously distinguish between a core and a cusp. } %

We conclude that measuring the proper motions of the stars   
{ of a sample with $N \gtrsim 2000$ objects can efficiently assess whether the mass density profile has a central core or a cusp}.  
In these analyses we adopted a null uncertainty $\sigma_v$ on the measures of the velocity components: non-null uncertainties would leave the estimated values unbiased, but they would increase the uncertainties on the estimated $\gamma$ { and $\lambda$}. The presence of a non-null uncertainty thus further increases the necessity of having a sufficiently large sample of stars with measured proper motions.

\begin{table}
 \caption{The estimated parameters of a model with a core,  $\gamma=0$, or a cusp, $\gamma=1$, and an isotropic velocity field, $\beta=0$. The input values of  $\log_{10}(\rho_{0}/ 10^7 \rm{M}_\odot \rm{kpc}^{-3})$, $\log_{10}(r_s/ \rm{kpc})$, { and of the  log-slope estimated at the half-light radius $\lambda(a)$ are $[2.28;-0.82; 1.70]$ for the model with $\gamma=0$  and $[0.73;-0.10;1.40]$ for the model with $\gamma=1$}. In the models labeled { ``-3D''},  
 all the three components of the velocity of each star are known. In the model labeled { ``-1D''},  
 only the line-of-sight components of the velocities are known, whereas the proper motion components are missing.}\label{tab:table_ccp_results}
  \begin{tabular}{lc|cc}
    \toprule
    { Parameters} & { Input Model} & { Core ($\gamma=0$)} & { Cusp ($\gamma=1$)} \\
    \midrule
    \multirow{4}{*}{$\log_{10}\biggl(\frac{\rho_{0}}{ 10^{7} \rm{M}_{\odot}/\rm{kpc}^{3}}\biggr)$} &  2000-3D & $2.17^{+0.29}_{-0.37} $ & $0.75^{+0.21}_{-0.24}$ \\[0.1cm]
    &  4000-3D & $1.97\pm0.29 $ & $0.73^{+0.19}_{-0.20}$ \\[0.1cm]
    & 6000-3D & $2.15^{+0.21}_{-0.23}$ & $0.90^{+0.13}_{-0.15}$   \\[0.1cm]
    & 6000-1D & $2.16\pm^{+0.32}_{-0.41}$  & $0.76^{+0.23}_{-0.29}$   \\[0.1cm]
     \midrule
        \multirow{4}{*}{$\log_{10}\biggl(\frac{r_{s}}{ \rm{kpc}}\biggr)$} &  2000-3D & $-0.79^{+0.14}_{-0.11} $ & $-0.11\pm0.11$ \\[0.1cm]
        & 4000-3D & $-0.70\pm 0.11$ & $-0.09\pm0.09$ \\[0.1cm]
        & 6000-3D & $-0.78^{+0.09}_{-0.08}$ & $-0.17\pm0.06$   \\[0.1cm]
    & 6000-1D & $-0.67^{+0.16}_{-0.12}$  & $0.0\pm0.13$   \\[0.1cm]
     \midrule
\multirow{4}{*}{$\gamma$} &    2000-3D & $0.19\pm0.40 $ & $0.97\pm0.11$ \\[0.1cm]
& 4000-3D & $0.33\pm^{+0.29}_{-0.36}$ & $0.98\pm0.10$ \\[0.1cm]
        & 6000-3D & $0.20\pm^{+0.26}_{-0.28}$ & $0.90\pm0.08$   \\[0.1cm]
    & 6000-1D & $0.53\pm0.36$  & $1.14\pm0.12$   \\[0.1cm] 
    \midrule
\multirow{4}{*}{$\lambda$} &    2000-3D & $1.73\pm0.04 $ & $1.38\pm0.04 $ \\[0.1cm]
& 4000-3D & $1.65\pm0.03 $ & $1.37\pm0.02 $ \\[0.1cm]
        & 6000-3D & $1.71\pm0.03 $ & $1.37\pm0.02 $  \\[0.1cm]
    & 6000-1D & $1.72\pm0.07$  & $1.45\pm0.06$   \\[0.1cm] 
    \bottomrule
  \end{tabular}
\end{table}

\section{Estimate of the mass profile of the dwarf}\label{sec:CCP&Mass}

\begin{figure*}
\begin{center}
\includegraphics[width=1.65\columnwidth, keepaspectratio]{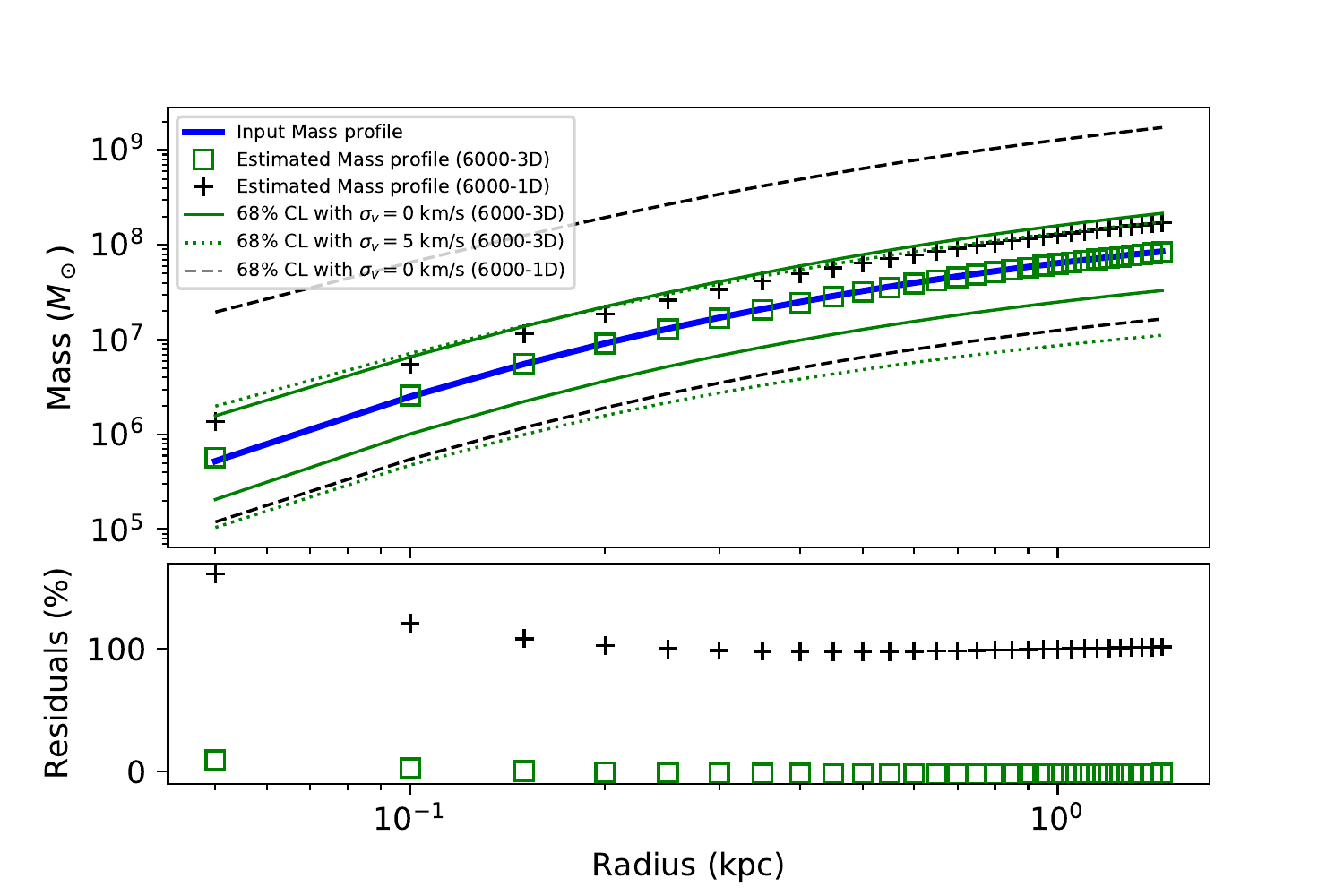}
\end{center}
\caption{{ Effect of the knowledge of the proper motions on the mass estimate. }{\it Upper panel}: Estimated cumulative mass profile of a dwarf model when the proper motions of 6000 stars are known (green open squares). The 68\% confidence region is limited by the green solid (dotted) lines when the velocity uncertainties are $\sigma_v=0$ (5) km~s$^{-1}$. The solid blue line is the correct cumulative mass profile. The crosses show the estimated cumulative mass profile when only the line-of-sight component of the velocities are known. The black dashed lines limit the 68\% confidence region in this case, when the velocity uncertainty is $\sigma_v=0$~km~s$^{-1}$. {\it Lower panel}: Residuals between the true (blue solid line) and the estimated (green open squares or black crosses) cumulative mass profiles. }
\label{fig:8}
\end{figure*}

We now explore how the proper motions can improve the estimate of the mass profile of our synthetic dwarfs. We consider the model where the density profile has a core. The input parameters $\rho_0$ and $r_s$  are listed in Table \ref{ParamsTable}, in the column with $\gamma=0$. The cumulative mass profile is provided by  Eq.~\eqref{eq:MassNFW} and shown by the blue solid line in Fig. \ref{fig:8}.

We now assume to observe this galaxy with the mock catalogue containing 6000 stars generated with the anisotropic velocity field $\beta=0$ and null uncertainty  $ \sigma_v=0$ km s$^{-1}$ on the measures of the velocity components. The last column of Table \ref{tab:table_mcmc_fit} lists the four parameters estimated with our MCMC approach. 
The green open squares in Fig. \ref{fig:8} show the estimated mass profile. Increasing the uncertainty on the velocity components to $ \sigma_v=5$~km~s$^{-1}$ returns an estimated mass profile that is indistinguishable from the mass profile estimated when the uncertainties are null.

The velocity uncertainty propagates into the uncertainty on the mass profile. We estimate the $1\sigma$ confidence regions of the estimated mass profiles by carrying out 1,000 Monte Carlo simulations: we randomly select the parameters of the DM profile, $(\rho_0, r_s, \gamma)$, from  the posterior distributions  whose means and amplitudes are listed in Table \ref{tab:table_mcmc_fit}. The result for $ \sigma_v=0$ km~s$^{-1}$ is shown in Fig. \ref{fig:8} as the area bounded by the green solid lines. Adding the observational error on the velocities broadens the bounded area (green dotted lines).  

The estimate of the mass profile substantially worsens when the proper motions are unavailable. The black crosses in Fig. \ref{fig:8} show the estimated mass profile when  only the line-of-sight component of the velocities are known. For this estimated profile we use the parameters listed in  Table \ref{tab:table_ccp_results} corresponding to the 6000-1D model.
In this case, the mass profile is overestimated by a factor 2. In addition, when the proper motions are unavailable, we recover the model parameters within $\sim 2\sigma$, rather than $\sim 1\sigma$, and the uncertainties are a factor $\sim 1.5- 2$ larger. This comparison quantifies the relevance of measuring the star proper motions to improve our knowledge of the DM distribution in dwarf galaxies.

This result is confirmed by the residuals between the true and the estimated mass profiles shown in the bottom panel of Fig. \ref{fig:8}. When the proper motions are available, the absolute values of the residuals range from a maximum of $\sim 9\%$ at $r\sim 0.05$ kpc to  a minimum of  $\sim 2.1\%$ at $r\sim 0.7$ kpc, as shown by the green squares in the bottom panel of Fig. \ref{fig:8}. When only the line-of-sight velocities are available, the absolute values of the residuals range from a maximum of $\sim 160\%$ at $r\sim 0.05$ kpc to  a minimum of $\sim 97\%$ at $r\sim 0.45$ kpc, as shown by the black crosses in the bottom panel of Fig. \ref{fig:8}. { We note that the residuals have a minimum close to the half-light radius of our model, in agreement with previous work \citep{Strigari2007,Wolf2010,Walker2009,Guerra2021}. 
The wide vertical scale of this panel inhibits the full appreciation of this minimum.} 

\begin{figure*}
\begin{center}
\includegraphics[width=1.65\columnwidth, keepaspectratio]{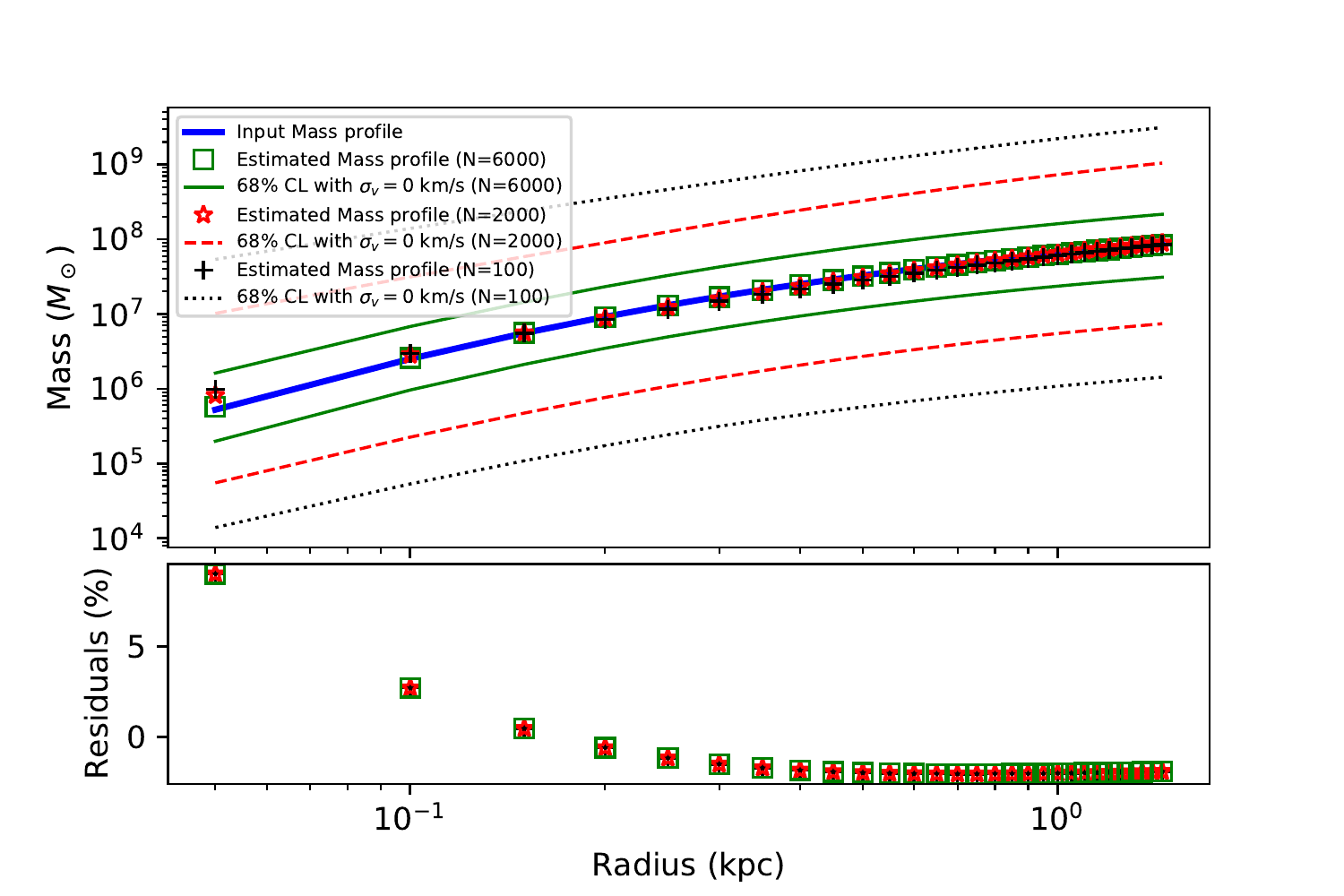}
\end{center}
\caption{{ Effect of the size of the star sample on the mass estimate. }{\it Upper panel:} Cumulative mass profile of a dwarf model estimated with a sample of 6000 (green open squares), 2000 (red stars), and 100 (black crosses) stars. The 68\% confidence regions are limited by the green solid, red dashed, and black dotted lines, respectively. The uncertainty on the velocity components is $\sigma_v=0$ km~s$^{-1}$. The blue solid line shows the true mass profile. {\it { Lower} panel}: Residuals between the true (blue solid line) and the estimated cumulative mass profiles. Symbols are as in the upper panel.}
\label{fig:9}
\end{figure*}

Finally,  Fig. \ref{fig:9} shows the impact of the size of the star sample on the estimated mass profile. We show the cumulative mass profile estimated with a sample of 6000, 2000, and 100 stars with known proper motions and their corresponding 68\% confidence regions: increasing the sample size narrows the 68\% confidence region, whereas the residuals remain unaffected. 

\section{Conclusions} \label{sec:discussion}

We determine the minimum number of stars and the minimum uncertainty on their measured proper motions that are necessary to distinguish  between a cusp and a core in the dark matter density profile of a dwarf galaxy, based on kinematic information alone. 
We created a set of astrometric mock catalogues by adopting the dark matter parameters of an extended Navarro-Frenk-White model and  the Plummer model of the stellar number density distribution of the Draco galaxy \citep{Walker2009}, which is a possible target of future space-borne astrometric missions \citep{Malbet2016, Theia2017, Malbet2019,Malbet2021}.

The accuracy of the estimates of the parameters mostly depend on the size of the sample, whereas velocity uncertainties $\le 5$ km~s$^{-1}$ only have a moderate impact on the parameter recovery. Our MCMC algorithm returns the dark matter parameters, including the velocity anisotropy parameter $\beta$, within $\sim 10\%$ of the true value and with a $1\sigma$ relative uncertainty $\lesssim 20\%$  for samples with $N\ge 2000$ stars. 
   
{ With a sample of $N=2000$ stars with measured proper motions, the measure of the log-slope $\lambda=-{\mathrm d}\ln\rho/{\mathrm d}\ln r$ of the density profile at the half-light radius distinguishes between a central core and a central cusp at more than $8\sigma$ in dwarfs described by our fiducial models \citep{Strigari2007,Guerra2021}. With the measure of the line-of-sight velocity components alone, the log-slope $\lambda$ can still distinguish the two models at more than $4\sigma$ with a sample of at least $N=6000$ stars. The validity of these results to dwarfs described by a more extended family of density profiles than we explored here remains to be investigated.
}

Proper motions enable the recovery of the cumulative mass profile of the dwarf with no bias with samples of stars as small as $N=100$ stars. Increasing the sample size reduces the uncertainty on the mass: with $N=6000$ stars the mass profile is on average 0.15 dex accurate.

In the future, we plan to investigate how the simplifying assumptions we adopted here can affect our results.
Indeed, we assumed (1) a constant anisotropy parameter $\beta$ rather than a radial dependent $\beta(r)$ \citep[e.g.,][]{Okoli_2015}, and (2) spherical symmetry for the synthetic dwarf, although real dwarfs appear elliptical on the sky
\citep[e.g.,][]{Irwin1995,Salomon2015}.  Moreover, we assumed a simplified scheme for the uncertainty on the star velocities: we  assumed the same error for the three components of the velocity, whereas, in reality, proper motions and line-of-sight velocities have substantially different uncertainties, as they depend on either astrometric or spectroscopic measurements. In addition, in our analysis we neglected the star brightness, which affects the uncertainties on the velocity components.  Finally, we assumed null errors on the distance of the dwarf from the observer, and of the projected distance of the stars from the dwarf centre.  More realistic mock catalogues will help to  better quantify the constraints on the dynamical properties of dwarfs and on the nature of dark matter that can be obtained { with 30 meter class telescopes \citep{Skidmore2015}},
with future astrometric space missions aimed at measuring the positions of celestial objects with precision of the order of the micro-arcsecond \citep{Malbet2016, Theia2017, Malbet2019,Malbet2021}, { or with a combination of the measures of these two classes of facilities \citep{Evslin2015}.} 

\section*{Acknowledgements}

We sincerely thank the active collaboration of Alistair Hodson during the initial development of this project. { We are in debt with a referee whose insightful comments crucially improved our results.} We acknowledge partial support from the INFN grant InDark and the Italian Ministry of Education, University and Research (MIUR) under the Departments of Excellence grant L.232/2016. 
IDM was supported by the grant``The Milky Way and Dwarf Weights with Space Scales'' funded by University of Torino and Compagnia di S. Paolo (UniTO-CSP).
IDM also acknowledges support from Ayuda  IJCI2018-036198-I  funded by  MCIN/AEI/  10.13039/501100011033  and  FSE  ``El FSE  invierte  en  tu  futuro''  o  ``Financiado  por  la  Unión  Europea   ``NextGenerationEU''/PRTR''. 
IDM is also supported by the project PGC2018-096038-B-I00  funded by the Spanish ``Ministerio de Ciencia e Innovación'' and FEDER ``A way of making Europe'', and by the project SA096P20 Junta de Castilla y León. 
This research has made use of NASA’s Astrophysics Data System Bibliographic Services.

\subsection*{Data Availability Statement}
No new data were generated or analysed in support of this research.


\vspace{1cm}

\bibliographystyle{mnras}
\bibliography{theia_refs} 



\appendix

\section{Relations between spherical and cartesian coordinate systems}\label{velocitymoments}

For completeness, we remind the formulae used to transform, in the reference frame centered on the centre of mass of the dwarf galaxy,  the mean velocity $\boldsymbol \mu$, the covariance matrix $\boldsymbol C$, and  the likelihood function in Eq.~\eqref{Likelihood2}, presented in  Sect.\ref{sec:SamplingPhaseSpace}, from the {system of} spherical coordinates  {$(r,\theta,\phi)$} to the system { of} Cartesian coordinates $(x,y,z)$. The transformation equation is 
\begin{equation}
\left(
\begin{array}{c}
v_x \\
v_y \\
v_z
\end{array}
\right) = \hat{R}\left(
\begin{array}{c}
v_r \\
v_\theta \\
v_\phi
\end{array}
\right)\,,
\end{equation}
where the rotation matrix $\hat{R}$ is 
\begin{equation}
\hat{R} = \left(
\begin{array}{ccc}
 \cos \phi  \sin \theta  & \cos \theta  \cos \phi  & -\sin \phi \\
 \sin \theta \sin \phi & \cos \theta \sin \phi & \cos \phi \\
 \cos \theta & -\sin \theta & 0 \\
\end{array}
\right)\, .
\end{equation}

In the spherical coordinate system, for non-rotating spherically symmetric systems, the first moments of the distributions of the three velocity components ${\rm v}_r$, ${\rm v}_\phi$, and ${\rm v}_\theta$ are null. Similarly, all the mixed terms of the form $\overline{{\rm v}_r{\rm v}_\phi}$ are null. In the Cartesian coordinates, we thus have 
\begin{align}
\overline{{{v}}_{x}} &= 0 \\
\overline{{{v}}_{y}} &= 0 \\
\overline{{{v}}_{z}} &= 0\, ,   
\end{align}
and 
\begin{align}
\overline{{{v}}_{x}^{2}}  = \cos ^2\phi \left({\overline{{{v}}_{r}^{2}}} \sin ^2\theta+{\overline{{{v}}_{\phi}^{2}} } \cos ^2\theta\right)+{\overline{{{v}}_{\phi}^{2}} } \sin ^2\phi
\end{align}
\begin{align}
 \overline{{{v}}_{y}^{2}}   =& \sin ^2\phi \left({\overline{{{v}}_{r}^{2}}} \sin ^2\theta+{\overline{{{v}}_{\phi}^{2}} } \cos ^2\theta\right)+{\overline{{{v}}_{\phi}^{2}} } \cos ^2\phi
\end{align}
\begin{align}
  \overline{{{v}}_{z}^{2}}  = &  {\overline{{{v}}_{r}^{2}}} \cos ^2\theta+{\overline{{{v}}_{\phi}^{2}} } \sin ^2\theta
\end{align}
\begin{align}
\overline{{{v}}_{x}{\rm{v}}_{y}} =     ({\overline{{{v}}_{r}^{2}}}-{\overline{{{v}}_{\phi}^{2}} }) \cos \phi \cos \theta \sin \theta
\end{align}
\begin{align}
\overline{{{v}}_{x}{\rm{v}}_{z}} = &  \sin ^2\phi \biggl({\overline{{{v}}_{r}^{2}}} \sin ^2\theta+{\overline{{{v}}_{\phi}^{2}} } \cos ^2\theta\biggr)+ {\overline{{{v}}_{\phi}^{2}} } \cos ^2\phi
\end{align}
\begin{align}
{\overline{{{v}}_{y}{\rm{v}}_{z}}} = &   -\frac{1}{2} \biggl({\overline{{{v}}_{r}^{2}}}-{\overline{{{v}}_{\phi}^{2}} }\biggr) \sin (2 \theta ) \sin \phi\, .
\end{align}

\bsp	
\label{lastpage}
\end{document}